\documentclass[twoside,openright,titlepage,numbers=noenddot,headinclude,footinclude=true,%
                cleardoublepage=empty,BCOR=5mm,paper=a4,fontsize=11pt,american]{scrbook}

\PassOptionsToPackage{utf8}{inputenc}
	\usepackage{inputenc}

\PassOptionsToPackage{listings,pdfspacing,floatperchapter,a5paper,subfig,parts,dottedtoc}{classicthesis}    

\newcommand{\myTitle}{A Classic Thesis Style\xspace}

\newcommand{\myName}{André Miede\xspace}

\newcommand{\myFaculty}{Put data here\xspace}

\newcommand{\myUni}{Put data here\xspace}

\newcounter{dummy} 
\providecommand{\mLyX}{L\kern-.1667em\lower.25em\hbox{Y}\kern-.125emX\@}

	\usepackage{babel}                  

\usepackage{csquotes}
\usepackage[comma,authoryear]{natbib}

\PassOptionsToPackage{fleqn}{amsmath}       
    \usepackage{amsmath,amsfonts,amssymb}

\PassOptionsToPackage{T1}{fontenc} 
    \usepackage{fontenc}     
\usepackage{textcomp} 
\usepackage{scrhack} 
\usepackage{xspace} 
\usepackage{mparhack} 
\usepackage{fixltx2e} 
\PassOptionsToPackage{printonlyused,smaller}{acronym} 
    \usepackage{acronym} 
    
\usepackage{tabularx,longtable,booktabs} 
\usepackage{caption}
\usepackage{subfig}  

\usepackage{listings} 
\PassOptionsToPackage{pdftex,hyperfootnotes=false,pdfpagelabels}{hyperref}
    \usepackage{hyperref}  
\PassOptionsToPackage{pdftex}{graphicx}
    \usepackage{graphicx}

\hypersetup{%
    colorlinks=true, linktocpage=true, pdfstartpage=3, pdfstartview=FitV,%
    breaklinks=true, pdfpagemode=UseNone, pageanchor=true, pdfpagemode=UseOutlines,%
    plainpages=false, bookmarksnumbered, bookmarksopen=true, bookmarksopenlevel=1,%
    hypertexnames=true, pdfhighlight=/O,
    urlcolor=webbrown, linkcolor=RoyalBlue, citecolor=webgreen, 
    pdftitle={\myTitle},%
    pdfauthor={\textcopyright\ \myName, \myUni, \myFaculty},%
    pdfsubject={},%
    pdfkeywords={},%
    pdfcreator={pdfLaTeX},%
    pdfproducer={LaTeX with hyperref and classicthesis}%
}   

\makeatletter
\@ifpackageloaded{babel}%
    {%
       \addto\extrasamerican{%
                }%
       \addto\extrasngerman{%
                }%
    }{\relax}
\makeatother

\usepackage{classicthesis} 

\usepackage{xcolor}
\usepackage{textcomp}
\usepackage{url}
\usepackage{version}
\usepackage{array}
\usepackage{enumitem}
\usepackage{tikz}
\usetikzlibrary{shapes,arrows,patterns}

\if@twocolumn
  \PackageInfo{classicthesis}{A5 paper, Palatino or other}%
  \areaset[current]{576pt}{555pt}%
  \recalctypearea
\else 
  \PackageInfo{classicthesis}{A4 paper, Palatino or other}
  \areaset[current]{336pt}{750pt} 
  \recalctypearea
\fi

\titleformat{\chapter}[display]%
        {\relax}{\mbox{}\flushright\vspace*{-2.5\baselineskip}\color{Maroon!70}\chapterNumber\thechapter}{0pt}%
        {\vspace*{1.5\baselineskip}\raggedright\color{Maroon}\spacedallcaps}[\normalsize\vspace*{1.5\baselineskip}]
\titleformat{\section}%
        {\relax}{\color{Maroon}\textsc{\MakeTextLowercase{\thesection}}}{1em}%
        {\color{Maroon} \spacedlowsmallcaps}
\titleformat{\subsection}%
        {\relax}{\color{Maroon}\textsc{\MakeTextLowercase{\thesubsection}}}{1em}%
        {\color{Maroon}\normalsize\itshape}

\lehead{\color{black!65}\mbox{\llap{\small\thepage\kern2em}\headmark\hfil}}
\rohead{\color{black!65}\mbox{\hfil{\headmark}\rlap{\small\kern2em\thepage}}}

\newenvironment{compactItemize}{\begin{itemize}[topsep=1ex,itemsep=0.1ex]}{\end{itemize}}

\definecolor{comment}{RGB}{0,100,0} 
\definecolor{comment}{RGB}{150,150,150} 
\definecolor{string}{RGB}{150,0,0}  
\definecolor{keyword}{RGB}{0,0,150} 

\lstdefinestyle{matlab}{
	commentstyle=\color{comment},
	stringstyle=\color{string},
	keywordstyle=\color{keyword},
	basicstyle=\footnotesize\ttfamily,
	numbers=left,
	numberstyle=\tiny,
	numbersep=5pt,
	frame=lines,
	breaklines=true,
	prebreak=\raisebox{0ex}[0ex][0ex]{\ensuremath{\hookleftarrow}},
	showstringspaces=false,
	upquote=true,
	tabsize=2
}

\definecolor{deepblue}{rgb}{0,0,0.5}
\definecolor{deepred}{rgb}{0.6,0,0}
\definecolor{deepgreen}{rgb}{0,0.5,0}

\lstdefinestyle{python2}{
language=Python,
basicstyle=\ttm,
otherkeywords={self},             
keywordstyle=\ttb\color{deepblue},
emph={MyClass,__init__},          
emphstyle=\ttb\color{deepred},    
stringstyle=\color{deepgreen},
frame=tb,                         
showstringspaces=false            %
}

\definecolor{dkgreen}{rgb}{0,0.6,0}
\definecolor{gray}{rgb}{0.5,0.5,0.5}
\definecolor{mauve}{rgb}{0.58,0,0.82}

\lstdefinestyle{python3}{
  frame=tb,
  language=Python,
  aboveskip=3mm,
  belowskip=3mm,
  showstringspaces=false,
  columns=flexible,
  basicstyle={\small\ttfamily},
  numbers=none,
  numberstyle=\tiny\color{gray},
  keywordstyle=\color{blue},
  commentstyle=\color{dkgreen},
  stringstyle=\color{mauve},
  breaklines=true,
  breakatwhitespace=true,
  tabsize=3
}

\newcommand{\Naturals}{{\mathbb{N}}}

\newcommand{\Reals}{{\mathbb{R}}}

\newcommand{\Vect}[1]{{\boldsymbol #1}}
\def\@biblabel#1{[{\bfseries #1}]}

\newcommand{\nvar}[2]{%
    \newlength{#1}
    \setlength{#1}{#2}
}

\nvar{\dg}{0.3cm}
\def\dw{0.25}\def\dh{0.5}
\nvar{\ddx}{1.5cm}

\def\link{\draw [double distance=1.5mm, very thick] (0,0)--}
\def\joint{%
    \filldraw [fill=white] (0,0) circle (5pt);
    \fill[black] circle (2pt);
}
\def\grip{%
    \draw[ultra thick](0cm,\dg)--(0cm,-\dg);
    \fill (0cm, 0.5\dg)+(0cm,1.5pt) -- +(0.6\dg,0cm) -- +(0pt,-1.5pt);
    \fill (0cm, -0.5\dg)+(0cm,1.5pt) -- +(0.6\dg,0cm) -- +(0pt,-1.5pt);
}
\def\robotbase{%
    \draw[rounded corners=8pt] (-\dw,-\dh)-- (-\dw, 0) --
        (0,\dh)--(\dw,0)--(\dw,-\dh);
    \draw (-0.5,-\dh)-- (0.5,-\dh);
    \fill[pattern=north east lines] (-0.5,-1) rectangle (0.5,-\dh);
}

\newcommand{\angann}[2]{%
    \begin{scope}[red]
    \draw [dashed, red] (0,0) -- (1.2\ddx,0pt);
    \draw [->, shorten >=3.5pt] (\ddx,0pt) arc (0:#1:\ddx);
    \node at (#1/2-2:\ddx+8pt) {#2};
    \end{scope}
}

\newcommand{\lineann}[4][0.5]{%
    \begin{scope}[rotate=#2, blue,inner sep=2pt]
        \draw[dashed, blue!40] (0,0) -- +(0,#1)
            node [coordinate, near end] (a) {};
        \draw[dashed, blue!40] (#3,0) -- +(0,#1)
            node [coordinate, near end] (b) {};
        \draw[|<->|] (a) -- node[fill=white] {#4} (b);
    \end{scope}
}

\def\thetaone{30}
\def\Lone{2}
\def\thetatwo{30}
\def\Ltwo{2}

\newtheorem{remark}{Remark}
\newtheorem{remarks}{Remarks}

\newtheorem{proposition}{Proposition}
\newtheorem{definition}{Definition}
\newtheorem{definitions}{Definitions}

\graphicspath{{GraphicsImages/}}

\renewcommand{\cite}[1]{\citep{#1}}

\begin{document}
\pagenumbering{roman}
\thispagestyle{empty}
\begingroup
\raggedleft 
\vspace*{\baselineskip} 
{\textcolor{black!65}{\Large Hugues Mounier}}\\[0.167\textheight] 
{\LARGE\bfseries Differential flatness for}\\[0.5\baselineskip] 
{\textcolor{Maroon}{\Huge  Neuroscience \\[-0.5ex] population \\[0.1ex] dynamics}}\\[2\baselineskip] 
{\Large {A preliminary study}}\par 
\vfill 
{\large Version 1 -- November 2016}\par 
\vspace*{3\baselineskip} 
\endgroup

\newpage\mbox{\ }\newpage
\vspace*{55ex}
\noindent
{\color{Maroon}Hugues Mounier} \\
Laboratoire des Signaux et Systèmes \\
CentraleSupélec \\
3, rue Joliot Curie \\
91192 GIF sur YVETTE \\
e-mail: {\color{Maroon}\ttfamily hugues.mounier@l2s.centralesupelec.fr}

\pagestyle{scrheadings}
\begin{small}
\tableofcontents 
 \end{small}

\cleardoublepage\pagenumbering{arabic}
 

\section*{Notations}
Some recurrent notations, transforms and sigmoid functions we shall be using throughout 
this document are given in Appendix \ref{app:notations} p.~\pageref{app:notations}.

\section*{Introduction}
The various objectives one whishes to attain through a controlled dynamical system
almost always boil down to a system's behavior modification. One has at his disposal
so called control variables whose aim is to steer the system. The behavior modification
generally consist in tracking a prescribed trajectory, with stability.
Note that the first phase (open loop trajectory tracking) is a feedforward one, while
the second (with stability) is a feedback one.
A possible methodology for controlling a system is then decomposed in two steps:
\begin{enumerate}
 \item \label{item:openLoop}
 A so called open loop trajectory tracking, supposing the model perfect and the
   initial conditions perfectly known.
 \item A feedback stabilizing the system around the reference trajectories, to compensate
   for model mismatch, poorly known initial conditions and external perturbations.
\end{enumerate}
Simple and natural solutions to problem \ref{item:openLoop} are obtained through the
differential flatness property, a notion due to Michel Fliess, Jean L\'e\-vi\-ne, Philippe 
Martin and Pierre Rouchon \cite{fliess-levine-martin-rouchon}.
This property amounts to a parametrization of a dynamical system in terms of a so-called 
flat output: any variable in the system can be expressed through a function of the flat
output components and a finite number of its derivatives.
This parametrization yields expressions of all the system's variables without having to
integrate any differential equation, which ensures fast computations. The control is in
particular given through the flat output and its derivatives ; an inversion of the 
system control input to flat output is thus performed, without any integration.
This notion has been extended to infinite dimensional systems, governed by partial
differential equations \cite{cvMouWoi10Siam}.

The present document is devoted to structural properties of neural population dynamics
and especially their differential flatness.
Several applications of differential flatness in the present context can be envisioned, 
among which:
trajectory tracking, feedforward to feedback switching, cyclic character,
positivity and boundedness.

\cleardoublepage
\part{Neural mass models}

\chapter{Simple and common neural mass models}
The following models are quite simple models for neural populations
with lumped space parameters (see, e.g. \cite{dayanAbbott-MITPress-2005}, Chapter 7,
\cite{ermentroutTerman-Springer-2010}, Chapter 11). The case of distributed parameter models,
so-called neural field models, is considered in Section \ref{sec:neuralFieldModels}, 
p.~\pageref{sec:neuralFieldModels}.

\section{Scalar integrate and fire models}
These type of models are of the form:
\begin{align}
 C \dot \nu &= -g_L (\nu - \nu_L) + F (\nu) + I 
\end{align}
where $\nu$ is the membrane potential, determined with respect to the
resting potential of the cell, $\tau_m$ is the membrane time
constant, $F(\nu)$ is a spike generating current, and $I$ is the total
current elicited by synaptic inputs to the neuron.

Common types of models, each associated with a specific type of spike 
generating currents, are:
\begin{compactItemize}
 \item The leaky integrate and fire, corresponding to $F = 0$
 \item The quadratic integrate and fire (or theta neuron), corresponding to 
   \begin{align*}
      F (\nu) &= \dfrac{g_L}{2 \Delta_T}\, (\nu - \nu_T)^2 + g_L (\nu - \nu_T) - I_T 
   \end{align*}
 \item The exponential integrate and fire, corresponding to 
   \begin{align*}
      F (\nu) &= g_L \Delta_T e^{\frac{\nu - \nu_T}{\Delta_T}}
   \end{align*}
\end{compactItemize}
\section{Two variables integrate and fire models}
More general models adds a second variable coupled to the voltage (see
\cite{izhikevich-2010})
\begin{align*}
 \dot \nu &= F (\nu) - \mu + I \\
 \dot \mu &= a (b\nu - \mu)
\end{align*}
The function $F(\nu)$ describes the current–voltage characteristic of the
membrane potential near the threshold, and it typically looks
like a parabola \cite{izhikevich-2003,izhikevich-2004}: $F(\nu) = \nu^2$.
Other choices possible are 
\begin{align*}
  F(\nu) &= |\nu|^3, \quad  F(\nu) = \dfrac{1}{1 - \nu}, \quad F(\nu) = |\nu|_+^n - \nu 
\end{align*}
An exponential spike generating current has been considered in 
\cite{bretteGerstner-2005} leading to the so-called adaptive exponential 
integrate and fire: $F(\nu) = e^\nu - \nu$, and \cite{touboul-2009} suggested the quartic
model $F(\nu) = \nu^4 + 2a \nu$.

\section{Two variables integrate and fire Izhikevich's models}
Another class of models is found in \cite{izhikevich-2010}:
\begin{align*}
 \dot \nu &= F (\nu) - \mu (E - \nu) + I \\
 \dot \mu &= a (b\nu - \mu)
\end{align*}
where $\nu$ plays the role of a conductance and $E$ is its reverse potential, which
could be assumed to take values $\pm 1$ or $0$ after appropriate rescaling. 

\section{Vectorial integrate and fire models}
Two types of integrate and fire models can be derived (see, e.g. 
\cite{ermentroutTerman-Springer-2010}, Chapter 11):
\begin{align}
 \label{eq:vectIntegrateAndFireMembrane}
 \tau_m \dot{\Vect{\nu}} &= -\Vect{\nu} + W \Vect{F}(\Vect{\nu}) + \tilde{\Vect{I}}
\end{align}
where $\tau_m$ is the membrane time constant, and
\begin{align}
 \label{eq:vectIntegrateAndFireSynaptic}
 \tau_d \dot{\Vect{\rho}} &= -\Vect{\rho} + \Vect{F}(W \Vect{\rho} + \Vect{I})
\end{align}
where $\tau_d$ is the synaptic decay time.
\begin{remarks}
 \begin{enumerate}
  \item The choice of one of the models is based on time scale considerations 
    (see, e.g. \cite{ermentroutTerman-Springer-2010}, p.~335), where in
    \eqref{eq:vectIntegrateAndFireSynaptic} the temporal dynamics is dominated by
    the synaptic decay and in \eqref{eq:vectIntegrateAndFireSynaptic}, the membrane 
    time constant of the postsynaptic cell are small compared with the decay of 
    the synapse.
  \item Note that the above two models can be shown to be equivalent (when $\tau_d = \tau_m$
    $= \tau$) in the following
    sense (see \cite{millerFumarola-2012}. If $\rho$ is a solution of the membrane 
    model \eqref{eq:vectIntegrateAndFireSynaptic}, then $W \Vect{\rho} + \Vect{I}$ is
    a solution of \eqref{eq:vectIntegrateAndFireMembrane}. Indeed, setting 
    $\nu = W \Vect{\rho} + \Vect{I}$, one obtains
    \begin{align*}
      \hspace*{-2ex}
      \tau \dot{\Vect{\nu}} &= \tau W \dot{\Vect{\rho}} + \tau \dot{\Vect{I}} =
          W \big( -\Vect{\rho} + \Vect{F}(W \Vect{\rho} + \Vect{I}) \big) + \tau \dot{\Vect{I}} \\
          &= -(\Vect{\nu} - \Vect{I}) + W \Vect{F}(\Vect{\nu}) + \tau \dot{\Vect{I}} \\
          &= -\Vect{\nu} +  W \Vect{F}(\Vect{\nu}) + \tilde{\Vect{I}}
    \end{align*}
 \end{enumerate}
\end{remarks}

\section{Neural mass Wilson-Cowan E-I networks}
Consider the simplest form of network, a pair of mutually coupled
local populations of excitatory and inhibitory neurons, also called E-I network
(see, e.g. \cite{bressloff-Springer-2014}, Subsection 6.2, p.~238). This model
was originally developed by Wilson and Cowan (see, e.g. 
\cite{ermentroutTerman-Springer-2010}, Subsection 11.3, p.~344), and has the
form
\begin{subequations}\begin{align}
 \label{eq:neuralMassWilsonCowanOne}
 \tau_e \dot \nu_e &= -\nu_e + F_e (w_{ee}\, \nu_e - w_{ie}\, \nu_i + I_e) \\
 \label{eq:neuralMassWilsonCowanTwo}
 \tau_i \dot \nu_i &= -\nu_i + F_i (w_{ii}\, \nu_i - w_{ei}\, \nu_e + I_i) 
\end{align}\end{subequations}
where $\nu_e$ and $\nu_i$ are the proportion of excitatory and inhibitory cells firing per 
unit time, the activations are nonlinear functions (typically sigmoidal) $F_e$, $F_i$ of 
the presently active proportion of cells, $w_{*}$ are the strength of the connections.

The matrix form of the previous model is
 \begin{subequations}\begin{align}
  \label{eq:matrixNneuralMassWilsonCowanOne}
  \Vect{\tau}_e \dot{\Vect{\nu}}_e &= -\Vect{\nu}_e + \Vect{F}_e (W_{ee}\, \Vect{\nu}_e - 
                                      W_{ie}\, \Vect{\nu}_i + \Vect{I}_e) \\
  \label{eq:matrixNeuralMassWilsonCowanTwo}
  \Vect{\tau}_i \dot{\Vect{\nu}}_i &= -\Vect{\nu}_i + \Vect{F}_i (W_{ii}\, \Vect{\nu}_i - 
                                      W_{ei}\, \Vect{\nu}_e + \Vect{I}_i) 
 \end{align}\end{subequations}

\chapter{Differential flatness}

\section{Differential flatness notion}
\subsection{Dynamics and observation equations}
Consider a system given by the dynamics equation and the observation equation
\begin{subequations}
\label{eq:sysNonLin}
  \begin{align}
  \label{eq:dynamics}
  \dot{\Vect{x}} &= f (\Vect{x},\Vect{u}) &\qquad &\text{dynamics equation}\\
  \label{eq:observation}
  \Vect{y}_m &= h (\Vect{x}) &\qquad &\text{observation equation}   
  \end{align}
\end{subequations}
with $\Vect{x}(t) = (x_1(t), \ldots, x_n (t))$, the \emph{state}, or, in Karl Friston's terms
the \emph{hidden variables} (see, e.g. \cite{friston-2012}), i.e. the controlled variables, 
$\Vect{u}(t) = (u_1(t), \ldots, u_m (t))$, the \emph{control input}, functions enabling an 
action on the process (typically input current), and $\Vect{y}_m = (y_{m1}(t), \ldots, y_{mp} (t))$ 
the \emph{output}, measured functions enabling to sense the environment (quantities coming 
from sensors).

Note that the dynamics equations form an \emph{undetermined} system of differential equations, 
since the control functions $\Vect{u}(t)$ are not a priori determined.
Once the control variables are fixed (i.e.\,\,substituted with known functions of time), the system 
\eqref{eq:sysNonLin} becomes determined (i.e.\,\,can be integrated). The \emph{state} variables 
represent the instantaneous memory of the system: once the control variables have been determined, 
the knowledge of the state variables (at time $t$) enables to predict the  future state (at 
time $t+dt$).

Another formualtion is the following: the state of a dynamical system is a set of
physical quantities the specification of which (in the absence of external excitation)
completely determines the evolution of the system.
\subsection{Differential flatness definition}
The notion of differential flatness (see \cite{fliess-levine-martin-rouchon}) is a form 
of controllability for non linear dynamical systems which is especially well suited for trajcetory 
tracking problems. It amounts to a parametrization of the system without integration of any
differential equation. Although the mathematical property seems quite strong, it appears that
this notion is commonly encountered in practice (see, e.g. \cite{rouchonJAMCS,martin-rouchon-JNCF} 
for a catalog of differentially flat systems). We shall give below a definition for such systems and 
illustrate this through simple examples derived from the well known Wilson and Cowan's model. Some 
more details about this property is given in the appendices.
\begin{definition}
The system
\begin{align}
  \label{Eq:sysplat}
  \dot{\Vect{x}} &= f (\Vect{x},\Vect{u})
\end{align}
with $\Vect{x}(t) \in \Reals^n$ and $\Vect{u}(t) \in \Reals^m$ is
\emph{differentially flat} if there exists a set of variables,
called a \emph{flat output},
\begin{align}
  \label{Eq:sortieplate}
\Vect{y} &= h (\Vect{x}, \Vect{u}, \dot{\Vect{u}}, \ldots,
\Vect{u}^{(r)}), \qquad \Vect{y}(t) \in \Reals^m, r \in \Naturals
\end{align}
such that
\begin{subequations}\begin{align}
  \label{Eq:sortieplatedeux}
\Vect{x} &= A (\Vect{y}, \dot{\Vect{y}}, \ldots, \Vect{y}^{(\rho_x)})\\
 \Vect{u} &= B (\Vect{y}, \dot{\Vect{y}}, \ldots, \Vect{y}^{(\rho_u)} )
\end{align}\end{subequations}
with $q$ an integer, and such that the system equations
\begin{align*}
\hspace*{-4ex}
\dfrac{dA}{dt}(\Vect{y}, \ldots, \Vect{y}^{(q+1)})
&=
  f(A(\Vect{y}, \ldots, \Vect{y}^{(q)}), B(\Vect{y},
  \ldots, \Vect{y}^{(q+1)}))
\end{align*}
are identically satisfied.
\end{definition}

\subsection{Parametrization}
For any flat output given through a function of the form 
$t \in \Reals \rightarrow \Vect{y} (t)$, the trajectory of the system
$\Vect{x}(t), \Vect{u}(t)$ are given by:
\begin{subequations}\begin{align}
 \label{eq:flatnessStateParametrization}
 \Vect{x}(t) &= A (\Vect{y}(t), \dot{\Vect{y}}(t), \ldots, \Vect{y}^{(\rho_x)}(t))\\
 \label{eq:flatnessInputParametrization}
 \Vect{u}(t) &= B (\Vect{y}(t), \dot{\Vect{y}}(t), \ldots, \Vect{y}^{(\rho_u)}(t) )
\end{align}\end{subequations}
There is a one to one correspondance between the system trajectories and the ones
given by the flat output.

\subsection{A word of methodology}
The preceding notion will be used to obtain so called ``open
loop'' controls, that is control laws which will ensure the
tracking of the reference flat outputs when the \emph{model is
assumed to be perfect} and \emph{the state initial conditions are
assumed to be exactly known}. Since this is never the case in
practice, one needs some feedback schemes that will ensure
asymptotic convergence to zero of the tracking errors. Our
framework can thus be decomposed in two steps:
\begin{enumerate}
\item Design of the reference trajectory of the flat outputs;
    off-line computation of the open loop controls.
\item Inline computation of the complementary closed loop controls
    in order to stabilize the system around the reference trajectories.
\end{enumerate}

Why is this two step design better suited than a classical
stabilization scheme? The first step obtains a first order
solution to the tracking problem, while \emph{following the model}
instead of forcing it (like in a usual pure stabilization scheme).
The second step is a refinement one, and the error between the
actual values and the tracked references will be much smaller than
in the pure stabilization case.

\section{Differential flatness of simple neural mass models}
\subsection{Weakly coupled E-I networks}
Consider a Wilson-Cowan model where $w_{ie} \ll 1$. Hence \eqref{eq:neuralMassWilsonCowanOne}
reduces to
\begin{align*}
 \tau_e \dot \nu_e &= -\nu_e + F_e (w_{ee}\, \nu_e + I_e)
\end{align*}
Alternatively, one could also consider the other limit case where $w_{ei} \ll 1$
where \eqref{eq:neuralMassWilsonCowanTwo} reduces to
\begin{align*}
 \tau_i \dot \nu_i &= -\nu_i + F_i (w_{ii}\, \nu_i + I_i) 
\end{align*}
Whatever case we consider, we shall abbreviate it by the following highly simplified model
\begin{align}
 \label{eq:neuralMassWilsonCowanWeaklyCoupled}
 \tau \dot \nu &= -\nu + F (w\, \nu + I) 
\end{align}
where the subscript has been dropped for convenience.
This model, although simplistic, is considered here because of its 
simplicity for pedagogical purposes.
Set
\begin{align*}
 \phi &= F^{-1}
\end{align*}
where $F$ is a sigmoid function (see \ref{subsec:sigmoidFunctions}, 
p.~\pageref{subsec:sigmoidFunctions}).

\subsection{Differential flatness of a weakly coupled E-I network}
The model depicted by \eqref{eq:neuralMassWilsonCowanWeaklyCoupled} is
differentially flat, with $\nu$ as a flat output. Indeed, one has
\begin{align*}
 w \nu + I &= \phi (\tau \dot \nu + \nu)
\end{align*}
and the input $I$ is given by
\begin{align}
 \label{eq:inputParametrizationNeuralMassWilsonCowanWeaklyCoupled}
 I &= -w \nu + \phi (\tau \dot \nu + \nu)
\end{align}

\subsection{Differential flatness of Wilson Cowan's E-I network}
The E-I network equations 
\eqref{eq:neuralMassWilsonCowanOne}--\eqref{eq:neuralMassWilsonCowanTwo}
\begin{align*}
 \tau_e \dot \nu_e &= -\nu_e + F_e (w_{ee}\, \nu_e - w_{ie}\, \nu_i + I_e) \\
 \tau_i \dot \nu_i &= -\nu_i + F_i (w_{ii}\, \nu_i - w_{ei}\, \nu_e + I_i) 
\end{align*}
rewrite 
\begin{align*}
 w_{ee}\, \nu_e - w_{ie}\, \nu_i + I_e &= F_e^{-1} (\tau_e \dot \nu_e + \nu_e) \\
 w_{ii}\, \nu_i - w_{ei}\, \nu_e + I_i &= F_i^{-1} (\tau_i \dot \nu_i + \nu_i)
\end{align*}
This model is thus differentially flat with flat output $(\nu_e, \nu_i)$:
\begin{align*}
 I_e &= w_{ie}\, \nu_i - w_{ee}\, \nu_e + F_e^{-1} (\tau_e \dot \nu_e + \nu_e) \\
 I_i &= - w_{ei}\, \nu_e + w_{ii}\, \nu_i + F_i^{-1} (\tau_i \dot \nu_i + \nu_i)
\end{align*}

\subsection{Differential flatness of asymetric Wilson Cowan's E-I network}
Consider the E-I Wilson-Cowan equations
\eqref{eq:neuralMassWilsonCowanOne}--\eqref{eq:neuralMassWilsonCowanTwo} with
statically coupled external currents
\begin{align*}
 I_e &= (1+a) I, \quad I_i = (1-a) I
\end{align*}
with $a \in [-1, 1]$ an asymetry factor.
The model \eqref{eq:neuralMassWilsonCowanOne}--\eqref{eq:neuralMassWilsonCowanTwo}
then becomes
(see, e.g. \cite{ermentroutTerman-Springer-2010}, Section 11.3, p.~ 349)
\begin{align*}
 \tau_e \dot \nu_e &= -\nu_e + F_e ((1+a) I - w_{ie}\, \nu_i) \\
 \tau_i \dot \nu_i &= -\nu_i + F_i ((1-a) I - w_{ei}\, \nu_e) 
\end{align*}

In the asymetric case, i.e. $a=-1$ the preceding equations are
\begin{subequations}\begin{align}
 \label{eq:asymetricNeuralMassWilsonCowanOne}
 \tau_e \dot \nu_e &= -\nu_e + F_e (- w_{i}\, \nu_i) \\
 \label{eq:asymetricNeuralMassWilsonCowanTwo}
 \tau_i \dot \nu_i &= -\nu_i + F_i (2 I - w_{e}\, \nu_e) 
\end{align}\end{subequations}
Then, $\nu_e$ is a flat output. Indeed, one gets
\begin{align*}
 \nu_i &= -\dfrac{1}{w_i}\, F_e^{-1} (\tau_e \dot \nu_e + \nu_e)  \\
 I   &= \dfrac{1}{2}\, \left[ w_e\, \nu_e + F_i^{-1} (\tau_i \dot \nu_i + \nu_i) \right]
\end{align*}

\chapter[Diff. flatness applications \& extensions]{Differential flatness applications and extensions}

\section{Differential flatness applications}
A number of applications of differential flatness can be envisioned, among which:
\begin{compactItemize}
 \item \emph{Trajectory tracking}.
 \item \emph{Feedforward to feedback switching}.
 \item \emph{Cyclic character}.
 \item \emph{Positivity} \& \emph{boundedness}.
 \item \emph{Simultaneous synchronisation} \& \emph{tracking}. 
\end{compactItemize}

\subsection{Trajectory tracking}
\label{subsec:tracking}
\subsubsection*{Flatness and feedback linearization}
A characterization of flat systems that appears
very useful for stabilized trajectory tracking is
the following
\begin{proposition}
A system is flat if, and only if, it is linearizable
by endogenous feedback and change of coordinates.
\end{proposition}
A dynamic feedback is called endogenous if it does
not include any external dynamics. More precisely
\begin{definition}
Consider the dynamics $\dot{\Vect{x}} = f(\Vect{x}, \Vect{u})$. 
The feedback
\begin{align*}
  \Vect{u} &= \xi (\Vect{x}, \Vect{z}, \Vect{v}) \\
  \dot{\Vect{z}} &= \zeta (\Vect{x}, \Vect{z}, \Vect{v})
\end{align*}
(where $\Vect{v}$ is the new input) is called a \emph{dynamic 
endogenous feedback}\index{Feedback!Dynamic endogenous} if the original dynamcis
$\dot{\Vect{x}} = f(\Vect{x}, \Vect{u})$ is equivalent to 
the transformed one
\begin{align*}
  \dot{\Vect{x}} &= f(\Vect{x},\xi (\Vect{x}, \Vect{z}, \Vect{v})) \\
  \dot{\Vect{z}} &= \zeta (\Vect{x}, \Vect{z}, \Vect{v})
\end{align*}
Two systems are called \emph{equivalent} if there exists a
invertible transformation wich exchanges their 
trajectories. 
\end{definition}

A more restrictive notion is the one of static state feedback, as described
below.
\begin{definition}
Consider the dynamics $\dot{\Vect{x}} = f(\Vect{x}, \Vect{u})$. 
The feedback
\begin{align*}
  \Vect{u} &= \xi (\Vect{x}, \Vect{v}) 
\end{align*}
(where $\Vect{v}$ is the new input) is called a 
\emph{static feedback}\index{Feedback!Static} if the original
dynamcis $\dot{\Vect{x}} = f(\Vect{x}, \Vect{u})$ is transformed to 
\begin{align*}
  \dot{\Vect{x}} &= f(\Vect{x},\xi (\Vect{x}, \Vect{v}))
\end{align*} 
\end{definition}
See the Subsection \ref{subsecStaticFeedbackLinearization}, 
p.~\pageref{subsecStaticFeedbackLinearization} for a static state feedback 
linearization criterion.

\subsubsection*{Dynamical extension algorithm}
This procedure enables one to know if an m-uple $(y_1, \ldots, y_m)$
is a flat output or not. Meanwhile, we shall obtain a linearizing 
feedback.
\paragraph{Phase I}-- Gathering the so called weak brunovsky indices.
\begin{itemize}
 \item[1)] Differentiate $y_1$ until a combination of controls appears.
     Note $\kappa_1$ the number of successive differentiations $y_1^{(\kappa_1)} = f_1$
 \item[2)] Differentiate $y_2$ until a combination of controls (independent
       of the previous ones) appears. Note $\kappa_2$ the number of successive 
       differentiations $y_2^{(\kappa_2)} = f_2$ \\[2ex]
 \vdots
 \item[m)] Differentiate $y_m$ until a combination of controls (independent 
       of the previous ones) appears. Note $\kappa_m$ the number of successive 
       differentiations $y_m^{(\kappa_m)} = f_m$ 
\end{itemize}
\paragraph{Phase II}-- Deciding the flatness character.\\
Then, if $\kappa_1 + \cdots + \kappa_m = n$ ($n$ being the state dimension), the system
admits $(y_1, \ldots, y_m)$ as a flat output. If not, $(y_1, \ldots, y_m)$ isn't  a flat output.
\paragraph{Phase III}-- Obtaining the linearizing feedback.\\
The linearizing feedback is given by $f_1 = v_1, \ldots, f_m = v_m$.

\subsubsection*{Closed loop trajectory tracking}
The open loop control laws suppose that the model is perfect and
that the initial conditions are exactly known. Since this is never
the case in practice, we add corrective terms to the open loop
controls derived above in order to stabilize the system around the
reference trajectories.

More precisely, considering a flat dynamics $\dot{\Vect{x}} = f(\Vect{x}, 
\Vect{u})$, we want to derive a controller able to follow any
reference trajectory $t \mapsto \Vect{y}_r (t)$.
In order to compensate for model mismatch and poorly
known initial conditions, one has to complement  the open
loop (obtained through flatness) with a closed loop corrective term depending 
on the error $\Vect{y}(t) - \Vect{y}_r(t)$.

Knowing the dynamics is flat, with flat output $\Vect{y}$,
it can be transformed via endogenous feedback and 
coordinate change to a linear dynamics of the form
\begin{align*}
   y_1^{(\kappa_1)} &= v_1 \\
  &\ \: \vdots \\
  y_m^{(\kappa_m)} &= v_m 
\end{align*}
with the new input $(v_1, \ldots, v_m)$.
Then, the elementary tracking feedback
\begin{align*}
  v_i &= y_{ir}^{(\kappa_i)} - \sum_{j=0}^{\kappa_i - 1} k_{ij} (y_i^{(j)} - y_{ir}^{(j)}), 
         \qquad i = 1, \ldots, m \\
      &=  y_{ir}^{(\kappa_i)} - \sum_{j=0}^{\kappa_i - 1} k_{ij} e_i^{(j)}
\end{align*}
with appropriately chosen $k_{ij}$ gains renders the error dynamics 
asymtotically stable:
\begin{align*}
  e_i^{(\kappa_i)} = \sum_{j=0}^{\kappa_i - 1} -k_{ij} e_i^{(j)}, \qquad i = 1, \ldots, m
\end{align*}

\subsection{Feedforward to feedback switching}
\subsubsection*{Open and closed loop}
The so-called \emph{open loop} control $\Vect{u}_o$ is obtained through 
\eqref{eq:flatnessInputParametrization}
\begin{align*} 
 \Vect{u}(t) &= B (\Vect{y}(t), \dot{\Vect{y}}(t), \ldots, \Vect{y}^{(\rho_u)}(t) )
\end{align*}
by replacing $y$ with a sufficiently differentiable trajectory $y_r (t)$:
\begin{align*} 
 \Vect{u}_o (t) &= B (\Vect{y}_r (t), \dot{\Vect{y}_r}(t), \ldots, \Vect{y}_r^{(\rho_u)}(t) )
\end{align*}
The use of this control law would lead to the desired tracking behavior $y = y_r$
if the model \eqref{eq:dynamics} was perfect and if the initial
conditions on $y$ was precisely known. Since this is never the case in practice,
one has to use closed loop feedback laws, such as the ones elaborated in the previous
Subsection \ref{subsec:tracking}, p.~\pageref{subsec:tracking}.
The difference between open and closed loop control laws can be bounded by the tracking
error and its derivatives. The simple weakly coupled E-I network example is examined
in Subsection \ref{subsec:openClosedWeakEINetwork}, p.~\pageref{subsec:openClosedWeakEINetwork}.

\subsubsection*{Temporal switching from feedforward to feedback}
Consider the following control law
\begin{align}
 \hspace*{-2ex}
 u(t) &= (1 - \sigma (t-t_{sw})) u_o (t) + \sigma (t-t_{sw}) u_c (t)
\end{align}
with $\sigma$ a sigmoid function, for example of the form
\begin{align*}
 \sigma(t) &= \dfrac{1}{1 + e^{\frac{-t-\beta}{\alpha}}}, \qquad
 \sigma(t) = \dfrac{1 + \tanh (\alpha t)}{2}
\end{align*}
Thus, from $t=0$ to $t=t_{sw} - d$ for some $d > 0$, we have $u \approx u_o$, and from 
$t= t_{sw} + d$, $u \approx u_c$.
This is the kind of control human beings tend to adopt for example in gesture control.
When grasping a glass of water, the first part of the gesture is done in open loop, quickly
and inaccurately; the second part of it is done with visual feedback, much more slowly
but precisely.

\subsection{Cyclic character}
When the flat output is cyclic, i.e.
\begin{align*}
 (1 - \Delta_\tau) y (t) &= y(t) - y(t-\tau) = 0
\end{align*}
Then, all the variables wich are expressed as functions of $y(t)$, that is
all the variables when the system is flat, are also cyclic:
for a variable $z$ which is expressed as $z = C(y, \dot y, \ldots, y^{(\eta)})$
\begin{align}
  \label{cyclicVariable}
  \hspace*{-5ex}
  (1 - \Delta_\tau) z &= (1 - \Delta_\tau) C(y, \dot y, \ldots, y^{(\eta)}) \notag \\
   &= C((1 - \Delta_\tau)y, (1 - \Delta_\tau) \dot y, \ldots, (1 - \Delta_\tau) y^{(\eta)}) \notag \\
   &= 0
\end{align}
More generally, if the flat output satifies a difference equation:
\begin{align*}
 p(\Delta_\tau) y &= 0
\end{align*}
where $p$ is a polynomial, then any variable of the flat system with flat output $y$
also satifies the same difference equation.

\subsection{Positivity \& Boundedness}
The goal is here to specify the reference trajectory in order to enforce certain properties
for various system variables.
Two main cases can be considered: The one of cyclic reference trajectories and the one
of non cyclic ones.

\if@twocolumn
  \vfill\vfill
\else \fi

\subsubsection*{Cyclic reference trajectories}
The flat output reference trajectories being cyclic can be expressed through a Fourier
series
\begin{align}
 \forall i=1, \ldots, m, \quad y_i &= \sum_{n=1}^\infty \xi_{i,n}\, e^{\frac{2j\pi}{n}}
\end{align}
Since all variables are also cyclic (see \eqref{cyclicVariable}) they can also 
be expressed through a Fourier series
\begin{align}
 z &= \sum_{n=1}^\infty \zeta_{n}\, e^{\frac{2j\pi}{n}}
\end{align}
One then has some relations expressing the $\zeta_n$ through the $\xi_{i,n}$:
\begin{align*}
 \zeta_n &= \phi_n (\xi_{1,n}, \ldots, \xi_{m,n})
\end{align*}
And the positivity can be expressed through a sum of squares type formula
(see, e.g. \cite{dumitrescu-2007}).
Several matlab packages are available for finding sum of squares decompositions of 
real multivariate polynomials; the most popular ones are SOSTOOLS (see 
\url{http://www.cds.caltech.edu/sostools/}), YALMIP (see 
\url{http://users.isy.liu.se/johanl/yalmip/}, and especially \\
\if@twocolumn
  \url{http://users.isy.liu.se/johanl/yalmip/pmwiki.php?n=Examples.MoreSOS}) and 
\else
  \url{http://users.isy.liu.se/johanl/yalmip/pmwiki.php?n=Examples}\\
  \url{.MoreSOS}) and 
\fi
GloptiPoly (see 
\url{http://homepages.laas.fr/henrion/software/gloptipoly/}).

\subsubsection*{Non cyclic reference trajectories}
One option is to take in the flat output $\Vect{y} = (y_1, \ldots, y_m)$ all the components 
$y_i$s as polynomial splines.
If the firing rate function is taken to be of Naka Rushton type, then all inequalities will
boil down to expressions of the form
\begin{align*}
 P_i(\Vect{y}, \dot{\Vect{y}}, \ldots, \Vect{y}^{(\rho)}) > 0, \qquad i=1, \ldots, m
\end{align*}
where the $P_i$s are polynomials in their variables. Since the $y_i$s are polynomial splines,
$P_i(\Vect{y}, \ldots, \Vect{y}^{(\rho)})$ will be another polynomial spline. 
And any approximating polynomial spline is contained in the convex hull of its control points.
One then can choose the lowest of these to be positive, to ensure the above inequality to
be fullfilled.

\subsection{Simultaneous synchronisation \& tracking} 
One considers here two (or more generally $N$) oscillators coupled via their input:
\begin{align*}
 \dot x_1 &= f_1 (\Vect{x}, u) \\
 \dot x_2 &= f_2 (\Vect{x}, u)
\end{align*}
The flatness of this system ensures not only that synchronisation is possible,
but also that any periodic trajectory (of the flat output) may be tracked, which
is a much stronger result.

\section[Flatness appls. for neural mass E-I networks]{%
Differential flatness applications for simple neural mass E-I networks}
\subsection{Trajectory tracking for weakly coupled E-I networks}
Recall the weakly coupled E-I network \eqref{eq:neuralMassWilsonCowanWeaklyCoupled},
p.~\pageref{eq:neuralMassWilsonCowanWeaklyCoupled}:
\begin{align}
 \tau \dot \nu &= -\nu + F (w\, \nu + I) 
\end{align}

In a trajectory tracking, one chooses a reference trajcetory $\nu_r$ and one 
wants that $\lim_{t \rightarrow \infty} \nu = \nu_r$, or, what is the same 
\begin{align*}
  \lim_{t \rightarrow \infty} e_\nu &= 0, \quad \text{where} \quad e_\nu = \nu - \nu_r
\end{align*}
This behavior can be enforced through the following desired error dynamics
\begin{align}
 \label{eq:expErrorDynamics}
 \dot e_\nu &= - \lambda e_\nu
\end{align}
where $\lambda > 0$ is a user chosen gain ruling the tracking error convergence
speed.
In order to obtain the desired behavior \eqref{eq:expErrorDynamics}, one has to
set in \eqref{eq:neuralMassWilsonCowanWeaklyCoupled}:
\begin{align*}
  -\nu + F(w\, \nu + I_c) &= -\tau \lambda e_\nu + \tau \dot \nu_r
\end{align*}
where $I_c$ is the closed loop control law. The preceding equation can be rewritten
as
\begin{align*}
   w\, \nu + I_c &= \phi (\nu -\tau \lambda e_\nu + \tau \dot \nu_r )
\end{align*}
which yields the following closed loop tracking feedback law 
\begin{align}
   \label{eq:closedLoopControlNeuralMassWilsonCowanWeaklyCoupled}
   I_c &= -w\, \nu + \phi (\tau \dot \nu_r  + \nu -\tau \lambda e_\nu)
\end{align}
which ensures, through \eqref{eq:expErrorDynamics}, the tracking of the reference
trajectory $\nu_r$ for the system \eqref{eq:neuralMassWilsonCowanWeaklyCoupled}
with stability.

\begin{remark}
 The application of the preceding extension algorithm is quite trivial since
 the system is fairly simple:
 \begin{itemize}
  \item Gathering the so called weak brunovsky indices.\\
      The flat output $\nu$ is differentiated once in equation 
      \eqref{eq:neuralMassWilsonCowanWeaklyCoupled} where the control $I$ is already
      present, hence $\kappa_1 = 1$.
  \item Deciding the flatness character.\\
      Since the dimension of the state is $n = 1$, $\sum_i \kappa_i = \kappa_1 = n$ 
      and the system is flat with flat output $\nu$.
  \item Obtaining the linearizing feedback.\\
     The linearizing feedback is given by:
     \begin{align}
        \label{eq:linearizingFeedbackNeuralMassWilsonCowanWeaklyCoupled}
        \dfrac{1}{\tau}\, \left( -\nu + F (w\, \nu + I) \right) &= v
     \end{align}
 \end{itemize}
 This feedback transforms the dynamics \eqref{eq:neuralMassWilsonCowanWeaklyCoupled} into
 the following linear one:
 \begin{align*}
  \dot \nu &= v
 \end{align*}
 and the elementary tracking feedback is
 \begin{align}
   \label{eq:elementaryTrackingNeuralMassWilsonCowanWeaklyCoupled}
   v &= \dot \nu_r -\lambda e_\nu
 \end{align}
 Thus, the original tracking feedback law is obtained from 
 \eqref{eq:linearizingFeedbackNeuralMassWilsonCowanWeaklyCoupled} and 
 \eqref{eq:elementaryTrackingNeuralMassWilsonCowanWeaklyCoupled}:
 \begin{align*}
    I &= -w\, \nu + \phi (\tau \dot \nu_r  + \nu -\tau \lambda e_\nu) \\
  \end{align*}
\end{remark}

\subsection{Trajectory tracking for asymetric Wilson-Cowan's E-I networks}
Recalling the equations of the asymetric Wilson-Cowan's E-I network
\eqref{eq:asymetricNeuralMassWilsonCowanOne}--\eqref{eq:asymetricNeuralMassWilsonCowanTwo}
\begin{align*}
 \tau_e \dot \nu_e &= -\nu_e + F_e (- w_{i}\, \nu_i) \\
 \tau_i \dot \nu_i &= -\nu_i + F_i (2 I - w_{e}\, \nu_e) 
\end{align*}
and differentiating the first equation in $\nu_e$, we get the flat output dynamics
\begin{align}
 \hspace*{-4ex}
 \label{eq:asymetricNeuralMassWilsonCowanFlatOutputDynamics}
 \tau_e \ddot \nu_e &= -\dot\nu_e - w_{i}\, F'_e (- w_{i}\, \nu_i) \dot \nu_i \notag\\
 &= -\dot\nu_e + \dfrac{w_{i}}{\tau_i}\, F'_e (- w_{i}\, \nu_i) 
          \big( \nu_i - F_i (2 I - w_{e}\, \nu_e) \big)
\end{align}
The desired dynamics being
\begin{align*}
  \ddot e_{er} &=  - \lambda_e e_{er} -\mu_{e} \dot e_{er}, \qquad 
                   \text{where } e_{er} = \nu_e - \nu_{er}
\end{align*}
the right hand side of \eqref{eq:asymetricNeuralMassWilsonCowanFlatOutputDynamics}
is then taken to be
\begin{align*}
  \hspace*{-5ex}
  &\dot\nu_e + \dfrac{w_{i}}{\tau_i}\, F'_e (- w_{i}\, \nu_i) 
          \big( -\nu_i + F_i (2 I - w_{e}\, \nu_e) \big) = \\
  &\hspace*{25ex} \tau_e \big( \ddot \nu_{er} + \lambda_e e_{er} + \mu_{e} \dot e_{er} \big)
\end{align*}
Thus we get 
\begin{align*}
  \hspace*{-7ex}
          &-\nu_i + F_i (2 I - w_{e}\, \nu_e)  = \\
  &\hspace*{8ex} \dfrac{\tau_i \tau_e}{w_{i} F'_e (- w_{i}\, \nu_i)}\, 
          \big( \dfrac{1}{\tau_e}\, \dot \nu_e + \ddot \nu_{er} +  
          \lambda_e e_{er} + \mu_{e} \dot e_{er} \big)
\end{align*}
and the tracking control feedback loop is obtained as
\if@twocolumn
  \begin{align*}
          I  &= \dfrac{1}{2}\, \Big[
          w_{e}\, \nu_e + F_i^{-1} \Big( \nu_i +
          \dfrac{\tau_i \tau_e}{w_{i} F'_e (- w_{i}\, \nu_i)}\, 
          \Big( \dfrac{1}{\tau_e}\, \dot \nu_e + \\ 
          &\hspace*{25ex}
          \ddot \nu_{er} + 
          \lambda_e e_{er} + \mu_{e} \dot e_{er} \Big) \Big) \Big]
  \end{align*}
\else 
  \begin{align*}
  \hspace*{-5ex}
          I  &= \dfrac{1}{2}\, \Big[
          w_{e}\, \nu_e + F_i^{-1} \Big( \nu_i +
          \dfrac{\tau_i \tau_e}{w_{i} F'_e (- w_{i}\, \nu_i)}\, 
          \Big( \dfrac{1}{\tau_e}\, \dot \nu_e + 
          \ddot \nu_{er} + 
          \lambda_e e_{er} + \mu_{e} \dot e_{er} \Big) \Big) \Big]
  \end{align*}
\fi

\subsection{Feedforward to feedback switching}
\label{subsec:openClosedWeakEINetwork}
\subsubsection*{Open and closed loop}
The so-called open loop control law is obtained when replacing $\nu$ by the 
reference trajectory $\nu_r$ in 
\eqref{eq:inputParametrizationNeuralMassWilsonCowanWeaklyCoupled}:
\begin{align}
  \label{eq:openLoopControlNeuralMassWilsonCowanWeaklyCoupled}
  I_o &= -w\, \nu_r + \phi (\tau \dot \nu_r  + \nu_r)
\end{align}
The use of this control law would lead to the desired tracking behavior $\nu = \nu_r$
if the model \eqref{eq:neuralMassWilsonCowanWeaklyCoupled} was perfect and if the initial
conditions on $\nu$ was precisely known.

This type of law is typically used by the brain, after training, for quick movements
where the sensory system is bypassed. When the sensory system is used, the so-called closed loop
control law \eqref{eq:closedLoopControlNeuralMassWilsonCowanWeaklyCoupled} is applied.

The difference between $I_c$ and $I_o$ is 
\begin{align*}
  \hspace*{-4ex}
 I_o - I_c &= w\, e_\nu - \phi (\tau \dot \nu_r + \nu - \tau \lambda e_\nu) 
              + \phi (\tau \dot \nu_r + \nu_r) \\
            &=  w\, e_\nu - \phi \big(\tau \dot \nu_r + \nu_r + (1- \tau \lambda) e_\nu\big) 
              + \phi (\tau \dot \nu_r + \nu_r) 
\end{align*}
Then, supposing $\phi$ to be globally $\gamma$-lipschitz:
\begin{align*}
  \hspace*{-4ex}
 \big| \phi \big(\tau \dot \nu_r\! +\! \nu_r \! +\! (1\!-\! \tau \lambda) e_\nu\big) 
              - \phi (\tau \dot \nu_r + \nu_r) \big| \leqslant \gamma 
              | 1\! -\! \tau \lambda | | e_\nu |
\end{align*}
Hence the difference $I_c - I_o$ admits the following bound
\begin{align*}
 |I_c - I_o| \leqslant \big( \alpha + \gamma | 1 - \tau \lambda | \big) | e_\nu |
\end{align*}
Thus, if the tracking error is small, $I_c$ is in a neighborhhod of $I_o$.

\subsubsection*{Temporal switching from feedforward to feedback}
Consider the following control law
\begin{align}
  \hspace*{-3ex}
 I(t) &= \big(1 - \sigma (t-t_{sw})\big) I_o (t) + \sigma (t-t_{sw}) I_c (t)
\end{align}
with $\sigma$ a sigmoid function (see \ref{subsec:sigmoidFunctions}, 
p.~\pageref{subsec:sigmoidFunctions}).
Thus, from $t=0$ to $t=t_{sw} - d$ for some $d > 0$, we have $I \approx I_o$, and from 
$t= t_{sw} + d$, $I \approx I_c$.
This is the kind of control human beings tend to adopt for example in gesture control.
When grasping a glass of water, the first part of the gesture is done in open loop, quickly
and inaccurately; the second part of it is done with visual feedback, much more slowly
but precisely.
The expression can alternatively be expressed as:
\if@twocolumn
  \begin{align*}
   I &=  \big(1 - \sigma_{sw} \big) \phi (\tau \dot \nu_r + \nu_r) + \\
       &\hspace*{15ex} \sigma_{sw} \big( 
       -w e_\nu + \phi (\tau \dot \nu_r + \nu - \tau \lambda e_\nu) \big) \\
     &= \phi (\tau \dot \nu_r + \nu_r) + \sigma_{sw} \big( 
        -w e_\nu + \Delta \phi (\nu, \nu_r) \big)
  \end{align*}
\begin{align*}
\hspace*{-4ex}
 \text{where } \sigma_{sw} &= \sigma (t-t_{sw}) \quad \text{and} \\
 \Delta \phi (\nu, \nu_r) &=  \phi (\tau \dot \nu_r + \nu - \tau \lambda e_\nu) - 
     \phi (\tau \dot \nu_r + \nu_r) 
\end{align*}
\else
  \begin{align*}
   I &=  \big(1 - \sigma_{sw} \big) \phi (\tau \dot \nu_r + \nu_r) + 
      \sigma_{sw} \big( 
       -w e_\nu + \phi (\tau \dot \nu_r + \nu - \tau \lambda e_\nu) \big) \\
     &= \phi (\tau \dot \nu_r + \nu_r) + \sigma_{sw} \big( 
        -w e_\nu + \Delta \phi (\nu, \nu_r) \big)
  \end{align*}
where $\sigma_{sw} = \sigma (t-t_{sw})$ and
  \begin{align*}
 \Delta \phi (\nu, \nu_r) &=  \phi (\tau \dot \nu_r + \nu - \tau \lambda e_\nu) - 
     \phi (\tau \dot \nu_r + \nu_r) 
\end{align*}
\fi
  %


\section[Diff. flatness appls. for simplistic motor control]{%
Differential flatness applications for simplistic motor control}
\subsection{Two link arm model}
Consider a two link robot arm acting as a simplistic model of a human arm:
\begin{subequations}
\label{eq:twoLinkArmModel}
\begin{align}
 M_{11} \theta_1 + M_{12} \theta_2 + C_1 (\Vect{\theta}, \dot{\Vect{\theta}}) + G_1 (\Vect{\theta}) 
   &= T_1 \\
 M_{21} \theta_1 + M_{22} \theta_2 + C_2 (\Vect{\theta}, \dot{\Vect{\theta}}) + G_2 (\Vect{\theta}) 
   &= T_2
\end{align} 
\end{subequations}
where $\theta_1$ is the angle of the first arm, $\theta_2$ of the second, 
$\Vect{\theta} = (\theta_1, \theta_2)^T$, $M_{ij}$ are equivalent masses, $C_i$ are
the coriolis forces, $G_i$ are the gravity forces, and $T_i$ are the control torques.
The expressions for the $C_i$'s and the $G_i$'s are the following:
\begin{itemize}[leftmargin=3ex]
 \item The inertia expressions are
   \begin{subequations}
   \begin{align}
     M_{11} &= J_1 + J_2 + m_1 r_1^2 + m_2 (l_1^2 + r_1^2 + 2 l_1 r_2 \cos \theta_2 ) \\
   \hspace*{-12ex}
     M_{12} = M_{21} &= J_2 + m_2 (r_2^2 + l_1 r_2 \cos \theta_2 ) \\
     M_{22} &= J_2 + m_2 r_2^2
   \end{align}
   \end{subequations}
   where $J_i$ is the inertia of link $i$, $m_i$ its mass, $l_i$ its length, and $r_i$ the
   distance from the beginning of the link to its center of mass.
 \item The Coriolis terms are given by:
   \begin{subequations}\begin{align}
     C_1 &= -m_2 l_1 \theta_2^2 r_2 \sin \theta_2 - 2 m_2 l_1 \dot \theta_1 \dot \theta_2 r_2
        \sin \theta_2 \\
     C_2 &= m_2 l_1 \dot \theta_1^2 r_2 \sin \theta_2
   \end{align}\end{subequations}    
  \item And the gravity terms are
   \begin{subequations}\begin{align}
     G_1 &= (m_2 l_1 + m_1 r_1) g \sin \theta_1 + m_2 r_2 g \sin (\theta_1 + \theta_2) \\
     G_2 &= m_2 r_2 g \sin (\theta_1 + \theta_2)
   \end{align}\end{subequations}    
\end{itemize}
Equations \eqref{eq:twoLinkArmModel} can be rewritten in a vectorial form; to this 
purpose, set
\begin{align*}
 M &= \begin{pmatrix}
       M_{11} & M_{12} \\
       M_{21} & M_{22} 
      \end{pmatrix}, \quad \Vect{\theta} = \begin{pmatrix}
        \theta_1 \\ \theta_2
      \end{pmatrix} \\ 
 \Vect{C} &= \begin{pmatrix}
        C_1 \\ C_2
      \end{pmatrix}, \quad \Vect{G} = \begin{pmatrix}
        G_1 \\ G_2
      \end{pmatrix}, \quad  \Vect{T} = \begin{pmatrix}
        T_1 \\ T_2
      \end{pmatrix}
\end{align*}
Then, model \eqref{eq:twoLinkArmModel} becomes
\begin{align}
 \label{eq:twoLinkArmVectorialModel}
 M \ddot{\Vect{\theta}} + \Vect{C}(\Vect{\theta}, \dot{\Vect{\theta}}) + 
   \Vect{G} (\Vect{\theta}) &= \Vect{T} 
\end{align}

\begin{figure}[h!]
\begin{center}
\begin{tikzpicture}
    \robotbase
    \angann{\thetaone}{$\theta_1$}
    \lineann[0.7]{\thetaone}{\Lone}{$l_1$}
    \link(\thetaone:\Lone);
    \joint
    \begin{scope}[shift=(\thetaone:\Lone), rotate=\thetaone]
        \angann{\thetatwo}{$\theta_2$}
        \lineann[-1.5]{\thetatwo}{\Ltwo}{$l_2$}
        \link(\thetatwo:\Ltwo);
        \joint
        \begin{scope}[shift=(\thetatwo:\Ltwo), rotate=\thetatwo]
                \grip
        \end{scope}
    \end{scope}
\end{tikzpicture}
\caption{\label{fig:twoLinkArm}A two link robot arm.}
\end{center}
\end{figure}
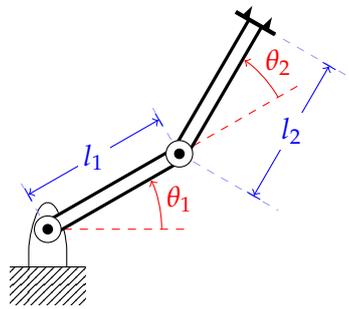

\subsection{Differential flatness and open loop control of the two link arm}
%
%
\begin{remark}
\label{rem:twoLinkArmFlatnessInTheta}
Model \eqref{eq:twoLinkArmVectorialModel} is differentially flat, with $\Vect{\theta}$
as a flat output. Indeed, the inputs $\Vect{T}$ are directly expressed in terms
of $\Vect{\theta}$ and its derivatives:
\begin{align*}
 \Vect{T}  &= M \ddot{\Vect{\theta}} + \Vect{C}(\Vect{\theta}, \dot{\Vect{\theta}}) + 
   \Vect{G} (\Vect{\theta})
\end{align*}
and the open loop control for a trajectory $\theta_{1r}, \theta_{2r}$ given by
\begin{align*}
 \Vect{T}_r  &= M \ddot{\Vect{\theta}}_r + \Vect{C}(\Vect{\theta}_r, \dot{\Vect{\theta}}_r) + 
   \Vect{G} (\Vect{\theta}_r)
\end{align*}
\end{remark}

Knowing that the desired trajectory is generally not given in terms of $\theta_1, 
\theta_2$ but in terms of the end effector coordinates $h_x, h_y$, we have to
express the former in terms of the latter. The end effector (e.g. the wrist) coordinates
are given by:
\begin{subequations}
\label{eq:endEffectorCoordinates}
\begin{align}
 h_x &= l_1 \cos \theta_1 + l_2 \cos (\theta_1 + \theta_2) \\
 h_y &= l_1 \sin \theta_1 + l_2 \sin (\theta_1 + \theta_2) 
\end{align}
\end{subequations}
The inversion of these formulae is detailed in Appendix \ref{app:twoLinkInverseKinematics},
p.~\pageref{app:twoLinkInverseKinematics}. We shall here give the final expressions:
\if@twocolumn
\begin{subequations}
\label{eq:twoLinkInverseKinematicsFinal}
\begin{align}
 \hspace*{-4ex}
 \theta_1 &= \arctan \left( \dfrac{h_y}{h_x} \right) - \arctan \left( 
 \dfrac{l_2 \sin \theta_2}{l_1 + l_2 \cos \theta_2} \right)
\end{align}
\begin{align}
  \theta_2 &= \arctan \left(
        \pm \dfrac{\sqrt{1 - \bar{h}^2}}{\bar{h}}
      \right)
\end{align}
\begin{align}
  \bar{h} &= \dfrac{h_x^2 + h_y^2 - l_1^2 - l_2^2}{2 l_1 l_2} \notag
\end{align}
\end{subequations}
\else
\begin{subequations}
\label{eq:twoLinkInverseKinematicsFinal}
\begin{align}
 \hspace*{-4ex}
 \theta_1 &= \arctan \left( \dfrac{h_y}{h_x} \right) - \arctan \left( 
 \dfrac{l_2 \sin \theta_2}{l_1 + l_2 \cos \theta_2} \right) \\
  \theta_2 &= \arctan \left(
        \pm \dfrac{\sqrt{1 - \bar{h}^2}}{\bar{h}}
      \right) \\
  \bar{h} &= \dfrac{h_x^2 + h_y^2 - l_1^2 - l_2^2}{2 l_1 l_2} \notag
\end{align}
\end{subequations}
\fi

\subsection{End effector dynamics}
%
%
The dynamcis in $\Vect{\theta}$ is given by:
\begin{align}
 \label{eq:dynTwoLinkArmInTheta}
 \ddot{\Vect{\theta}} &= -M^{-1} (\Vect{C} + \Vect{G}) + M^{-1} \Vect{T} 
\end{align}
and the dynamcis in the end effector, i.e. in $h_x$, $h_y$ is obtained through
a double differentiation of \eqref{eq:endEffectorCoordinates}. A first differentiation
yields
\begin{alignat*}{3}
 \dot h_x &= &-&l_1 \dot \theta_1 \sin \theta_1 - l_2 (\dot \theta_1 + \dot \theta_2) 
              \sin (\theta_1 + \theta_2) \\
 \dot h_y &= &&l_1 \dot \theta_1 \cos \theta_1 + l_2 (\dot \theta_1 + \dot \theta_2) 
              \cos (\theta_1 + \theta_2)
\end{alignat*}
And then
\begin{subequations}
\begin{align}
 \ddot h_x &= -h_y \ddot \theta_1 - l_2 \sin (\theta_1 + \theta_2) \ddot \theta_2 - 
               \phi_x (\Vect{\theta}, \dot{\Vect{\theta}}) \\
 \ddot h_y &=  h_x \ddot \theta_1 + l_2 \cos (\theta_1 + \theta_2) \ddot \theta_2 - 
               \phi_y (\Vect{\theta}, \dot{\Vect{\theta}})
\end{align} 
\end{subequations}
with
\begin{align*}
 \phi_x (\Vect{\theta}, \dot{\Vect{\theta}}) &= l_1 \dot \theta_1^2 \cos \theta_1 +
     l_2 (\dot \theta_1 + \dot \theta_2)^2 \cos (\theta_1 + \theta_2) \\
 \phi_y (\Vect{\theta}, \dot{\Vect{\theta}}) &= l_1 \dot \theta_1^2 \sin \theta_1 +
     l_2 (\dot \theta_1 + \dot \theta_2)^2 \sin (\theta_1 + \theta_2) 
\end{align*}
Thus, one gets
\begin{align*}
 \begin{pmatrix}
   \ddot h_x \\ \ddot h_y
 \end{pmatrix} &=
 \begin{pmatrix}
  -h_y  &  -l_2 \sin (\theta_1 + \theta_2) \\ 
   \ \ h_x  &   \ \ l_2 \cos (\theta_1 + \theta_2)
 \end{pmatrix}
 \begin{pmatrix}
  \ddot \theta_1 \\ \ddot \theta_2
 \end{pmatrix} -
 \begin{pmatrix}
  \phi_x \\ \phi_y
 \end{pmatrix}
\end{align*}
Or, in other terms
\begin{align*}
   \ddot{\Vect{h}} &=
  H\, \ddot{\Vect{\theta}} -
  {\Vect{\phi}}
\end{align*}
With the following notations 
\begin{align}
  \label{eq:MatrixNotationsForWristDynamics}
  \hspace*{-3ex}
  H &= \begin{pmatrix}
  -h_y  &  -l_2 \sin (\theta_1 + \theta_2) \\ 
   \ \ h_x  &   \ \ l_2 \cos (\theta_1 + \theta_2)
  \end{pmatrix}, \quad
   \Vect{h} = \begin{pmatrix}
   h_x \\ h_y
 \end{pmatrix}
\end{align}

And, using \eqref{eq:dynTwoLinkArmInTheta}, one gets the dynamics in the end effector
$\Vect{h}$:
\begin{align}
 \label{eq:dynTwoLinkArmInH}
   \ddot{\Vect{h}} &=
  -H\,M^{-1} \big(\Vect{C} + \Vect{G} - \Vect{T} \big) -
  {\Vect{\phi}}
\end{align}
\subsection{Trajectory tracking of the two link arm}
The system \eqref{eq:dynTwoLinkArmInH} is differentially flat, with flat output
$h_x, h_y$. Indeed equations \eqref{eq:twoLinkInverseKinematicsFinal} yield the 
expressions of $\theta_1$ and $\theta_2$ in terms of $h_x, h_y$, and $\Vect{T}$ is given by:
\begin{align*}
 \Vect{T} &= \Vect{C} + \Vect{G} + M H^{-1} \big( \ddot{\Vect{h}} + {\Vect{\phi}} \big)
\end{align*}
Thus, considering a reference trajectory $h_{xr} (t)$, $h_{yr} (t)$, the so-called open loop
control $\Vect{T}_r$ is given by:
\begin{align}
 \label{eq:twoLinkArmOpenLoopControl}
  \Vect{T}_r &= \Vect{C}_r + \Vect{G}_r + M H_r^{-1} \big( \ddot{\Vect{h}}_r + {\Vect{\phi}}_r \big)
\end{align}
with
\begin{align*}
 \hspace*{-5ex}
 \Vect{C}_r &= \begin{pmatrix}
     -m_2 l_1 \theta_{2r}^2 r_2 \sin \theta_{2r} - 2 m_2 l_1 \dot \theta_{1r} \dot \theta_{2r} r_2
        \sin \theta_{2r} \\
     m_2 l_1 \dot \theta_{1r}^2 r_2 \sin \theta_{2r}                
               \end{pmatrix} 
\end{align*}
\begin{align*}
 \hspace*{-5ex}
 \Vect{G}_r &= \begin{pmatrix}
     (m_2 l_1 + m_1 r_1) g \sin \theta_{1r} + m_2 r_2 g \sin (\theta_{1r} + \theta_{2r}) \\
     m_2 r_2 g \sin (\theta_{1r} + \theta_{2r})
               \end{pmatrix} 
\end{align*}
\begin{align*}
 \hspace*{-5ex}
 H_r &= \begin{pmatrix}
       -h_{yr}  &  -l_2 \sin (\theta_{1r} + \theta_{2r}) \\ 
   \ \ h_{xr}  &   \ \ l_2 \cos (\theta_{1r} + \theta_{2r})
               \end{pmatrix} 
\end{align*}
\begin{align*} 
 \hspace*{-5ex}
 \Vect{\phi}_r &= \begin{pmatrix}
   l_1 \dot \theta_{1r}^2 \cos \theta_{1r} +
     l_2 (\dot \theta_{1r} + \dot \theta_{2r})^2 \cos (\theta_{1r} + \theta_{2r}) \\
 l_1 \dot \theta_{1r}^2 \sin \theta_{1r} +
     l_2 (\dot \theta_{1r} + \dot \theta_{2r})^2 \sin (\theta_{1r} + \theta_{2r})     
               \end{pmatrix} 
\end{align*}
\begin{align*}  
 \hspace*{-5ex}
 \theta_{1r} &= \arctan \left( \dfrac{h_{yr}}{h_{xr}} \right) - \arctan \left( 
 \dfrac{l_2 \sin \theta_{2r}}{l_1 + l_2 \cos \theta_{2r}} \right) \\
 \hspace*{-5ex}
  \theta_{2r} &= \arctan \left(
        \pm \dfrac{\sqrt{1 - \bar{h}_r^2}}{\bar{h}_r}
      \right) \\
  \bar{h}_r &= \dfrac{h_{xr}^2 + h_{yr}^2 - l_1^2 - l_2^2}{2 l_1 l_2} \notag
\end{align*}

Then, the feedback control law ensuring tracking of the reference trajectory
$h_{xr} (t)$, $h_{yr} (t)$ is given by:
\begin{align}
 \label{eq:twoLinkArmTrackingFeeback}
  \hspace*{-4ex}
 \Vect{T}  &= \Vect{C} + \Vect{G} + M H^{-1}\!\!\: \left( {\Vect{\phi}}\!\!\: +\!\!\:
   \ddot{\Vect{h}}_r \!\!\: -\!\!\: {\Lambda}_{0}^h \Vect{e}_h \!\: -\!\: 
   {\Lambda}_{1}^h \dot{\Vect{e}}_h
   \right)
\end{align}
with
\begin{align*}
 {\Lambda}_{0}^h &= \begin{pmatrix}
                    \lambda_{00}^h & 0 \\
                    0 & \lambda_{01}^h
                   \end{pmatrix}, \quad 
 {\Lambda}_{1}^h = \begin{pmatrix}
                    \lambda_{10}^h & 0 \\
                    0 & \lambda_{11}^h
                   \end{pmatrix}
\end{align*}
where the $\lambda_{ijh}$ are suitably chosen reals such that the closed loop
error equation in $\Vect{e}_h$ is exponentially stable (it is thus sifficient to choose
these as strictly positive reals).

Note that the difference between the previous tracking control law and the 
feedforward one given in \eqref{eq:twoLinkArmOpenLoopControl} is of the form:
\begin{align}
 \label{eq:twoLinkArmClosedAndOpenLoopDifference}
 \Vect{T} - \Vect{T}_r &= \Vect{C} - \Vect{C}_r + \Vect{G} - \Vect{G}_r + \notag \\
   &\hspace*{4ex} M H^{-1} \left( {\Vect{\phi}} + \ddot{\Vect{h}}_r \right) 
   - M H_r^{-1} \big( {\Vect{\phi}}_r + \ddot{\Vect{h}}_r \big) - \notag \\
   &\hspace*{4ex} M H^{-1} \left(  \Vect{\lambda}_h^T \Vect{e}_h + 
     \Vect{\mu}_h^T \dot{\Vect{e}}_h
   \right)
\end{align}
which tends to zero when $\Vect{e}_h$ itself tends to zero.

In \eqref{eq:twoLinkArmTrackingFeeback}, one needs to compute $H^{-1}$ (the matrix $H$ being
defined in \eqref{eq:MatrixNotationsForWristDynamics}), which requires
the determinant $\Delta_H$
\begin{align*}
 \Delta_H &= l_2 \big( h_x \sin (\theta_1 + \theta_2) - h_y \cos (\theta_1 + \theta_2) \big)
\end{align*}
to be non zero. From \eqref{eq:endEffectorCoordinates}, we get 
\begin{align*}
 \Delta_H &= l_1 \big( h_x \sin \theta_1  - h_y \cos \theta_1 \big)
\end{align*}
Thus, when $\Delta_H = 0$, we get 
\begin{align*}
 \tan (\theta_1 + \theta_2) &= \tan \theta_1
\end{align*}
Or, what is the same
\begin{align*}
 \theta_2 &= 0, \quad \text{or} \quad \theta_2 = \pi
\end{align*}
The first case yields the following end effector coordinates:
\begin{align*}
 h_x &= (l_1 + l_2) \cos \theta_1 \\
 h_y &= (l_1 + l_2) \sin \theta_1
\end{align*}
Thus, the end effector with coordinates $h_x$, $h_y$ remains on
a circle centered at the origin and with radius $l_1 + l_2$, which corresponds to the arm being
fully extended. The second case ($\theta_2 = \pi$) yields the end effector coordinates:
\begin{align*}
 h_x &= (l_1 - l_2) \cos \theta_1 \\
 h_y &= (l_1 - l_2) \sin \theta_1
\end{align*}
and the end effector with coordinates $h_x$, $h_y$ remains on
a circle centered at the origin and with radius $l_1 - l_2$, wich corresponds to the arm 
fully folded.

When designing a reference trajectory, we shall avoid these two cases. Let us consider
\begin{subequations}
\begin{align}
 \label{eq:twoLinkArmEndEffectorRefTraj}
 h_{yr} (t) &= \dfrac{h_{yf} - h_{yi}}{2}\, \left[ 1 + 
               \tanh \Big( \gamma (h_{xr}(t) - h_{x0}) \Big)\right] \\
 h_{xr} (t) &= \dfrac{(h_{xf} - h_{xi}) t}{T} + h_{xi}
\end{align} 
\end{subequations}
for $t \in [0,T]$, and for example:
\begin{subequations}
\begin{align}
 h_{xi} &= 0.8 (l_1 + l_2), \quad h_{xf} = 0 \\
 h_{yi} &= l_1 + 0.1 l_2, \quad h_{yf} = -0.1 l_1
\end{align} 
\end{subequations}

\begin{figure}[!h]
\subfloat[End effector tracking: $h_y$ versus $h_x$.]{%
  \includegraphics[width=.5\linewidth,clip,trim=20 180 40 250]{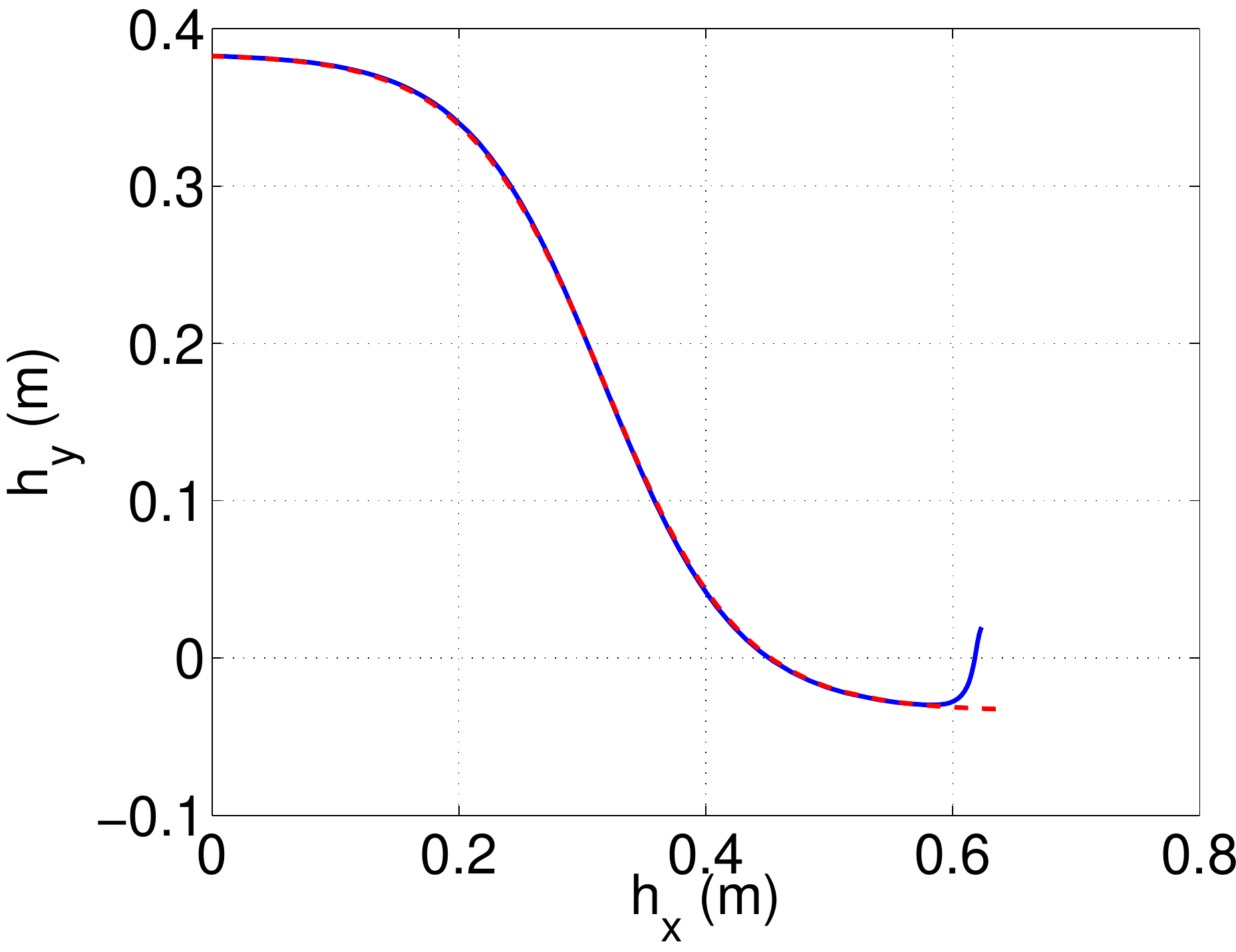}
  \label{fig:twoLinkArmHxHy} 
}
\subfloat[Animated tracking of the two link arm.]{%
  \includegraphics[width=.5\linewidth,clip,trim=25 180 40 250]{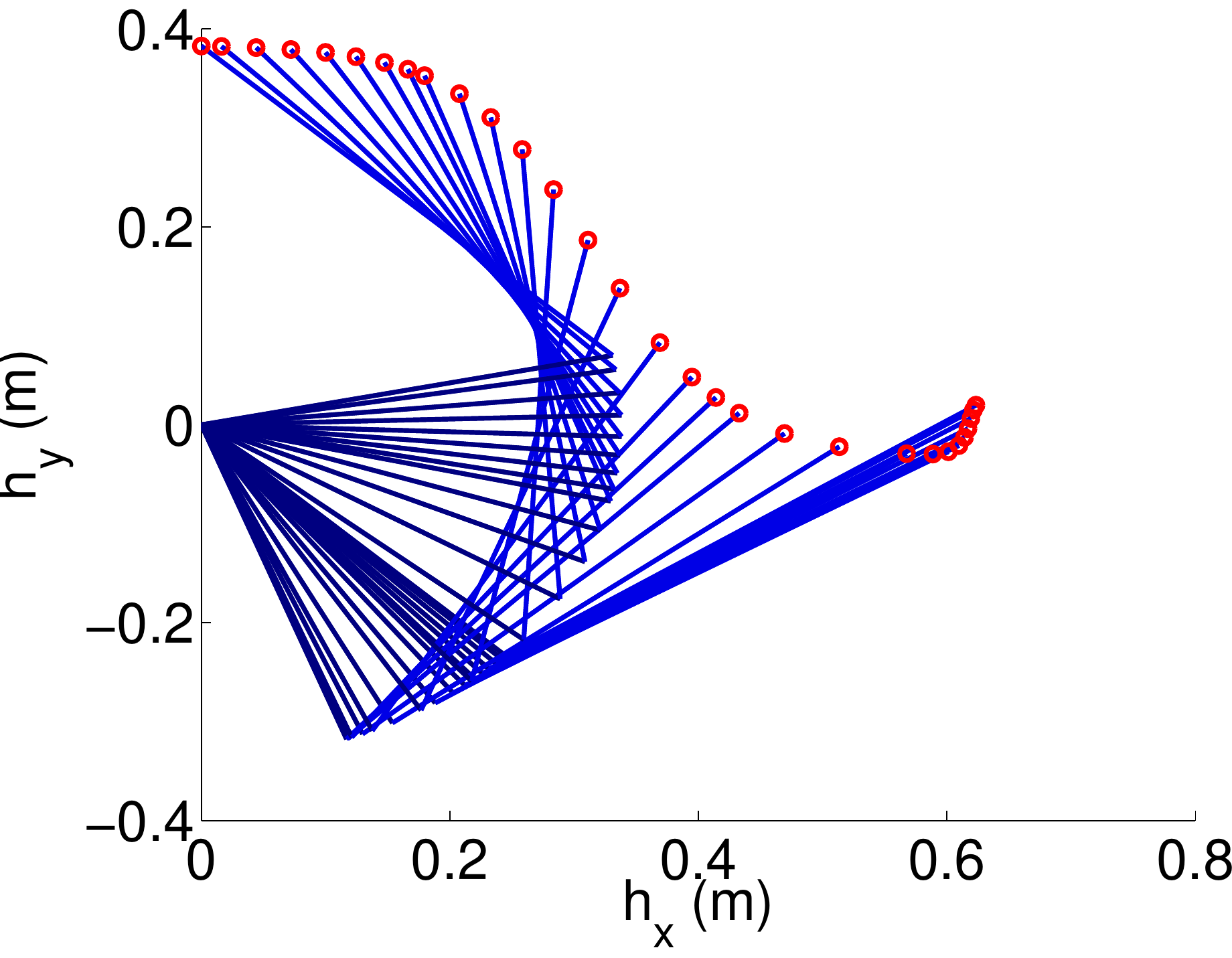}
  \label{fig:twoLinkArmAnimation}
}
\caption{Trajectory tracking of a two link robot arm. In red and dashed the reference
  trajectory; in blue and solid, the actual (simulated) trajectory.}
\label{fig:twoLinkArmTracking}
\end{figure}

\begin{figure}[!h]
\subfloat[Tracking feedback control $T_1$.]{%
  \includegraphics[width=.5\linewidth,clip,trim=20 180 40 250]{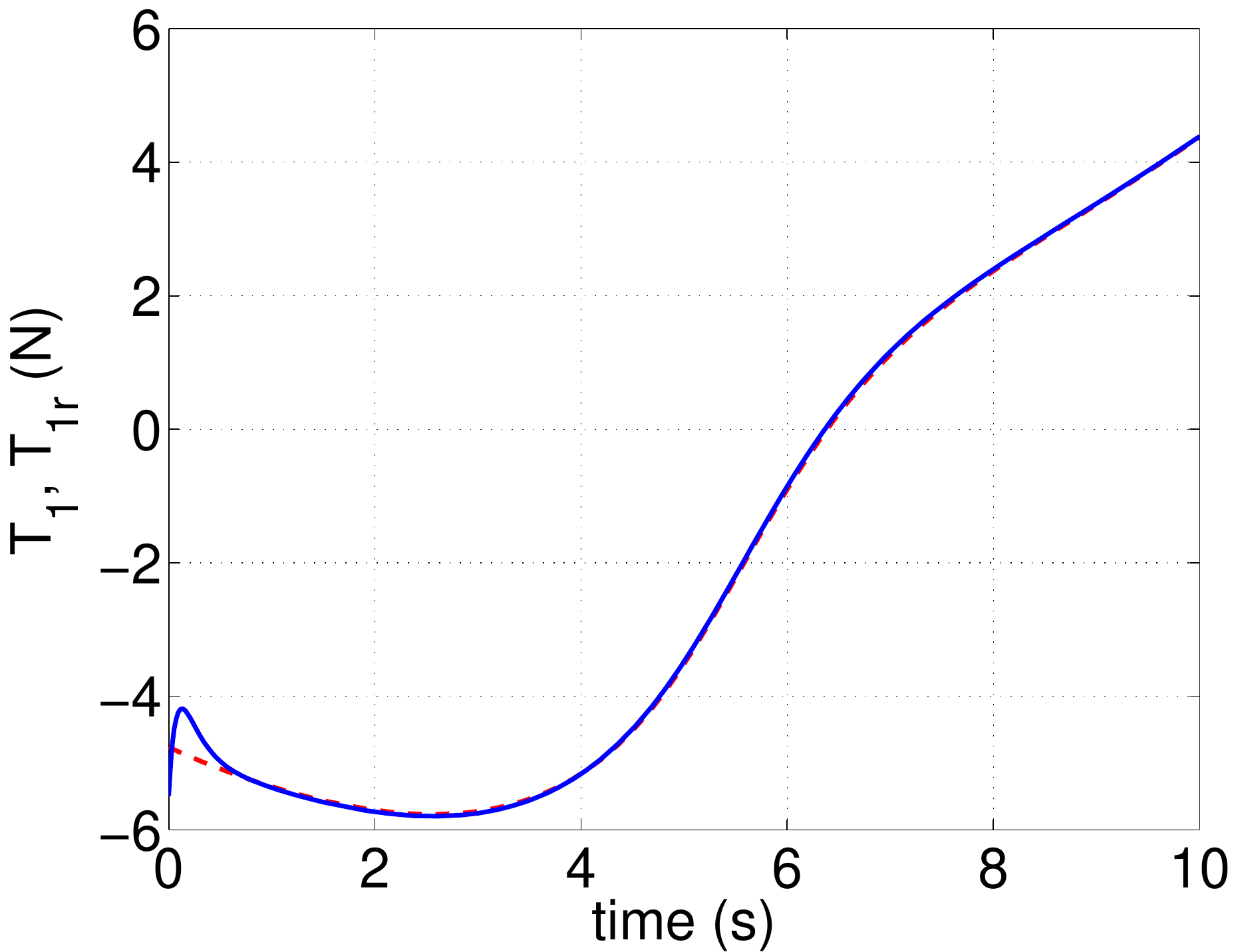}
  \label{fig:twoLinkArmCtrl1} 
}
\subfloat[Tracking feedback control $T_2$.]{%
  \includegraphics[width=.5\linewidth,clip,trim=20 180 40 250]{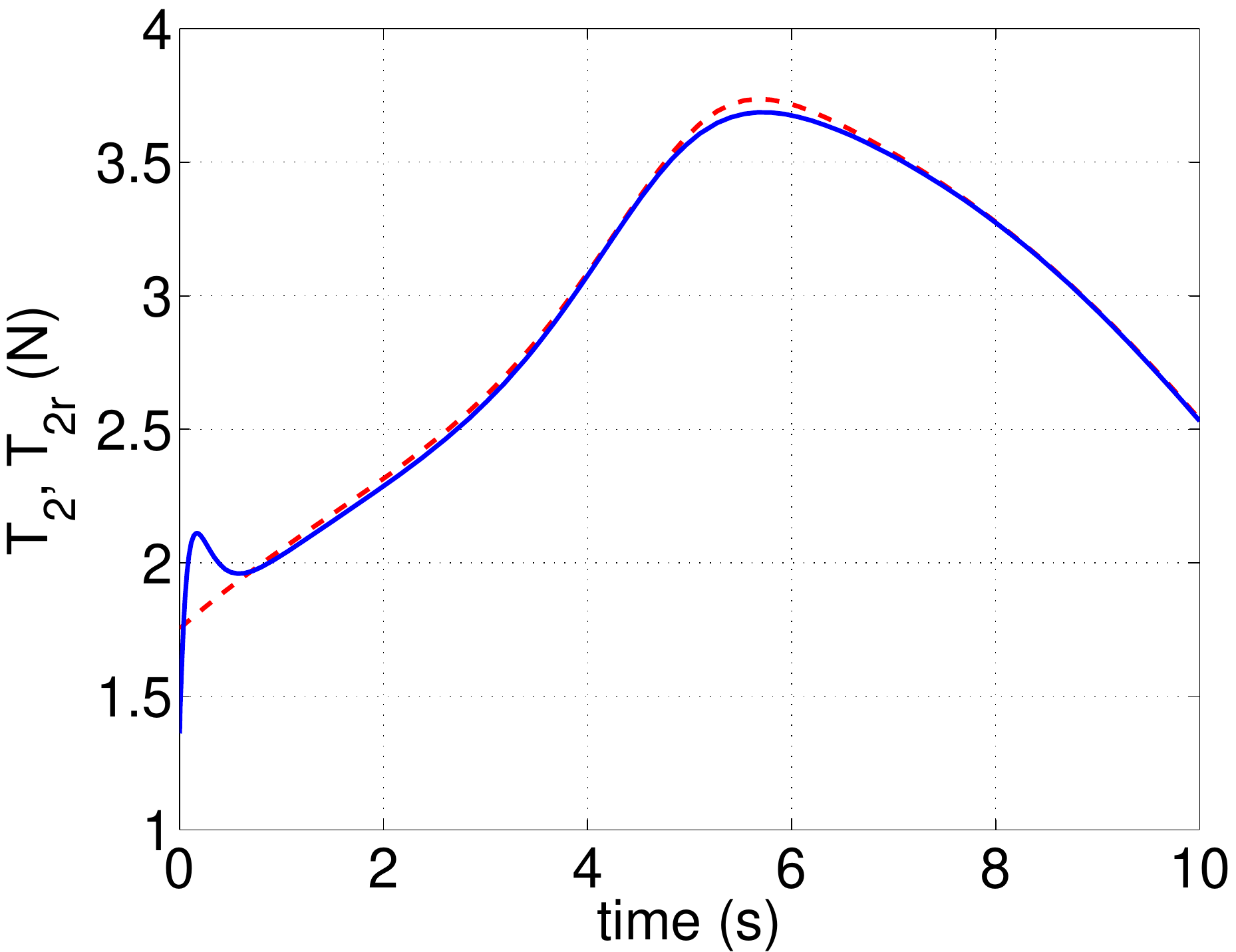}
  \label{fig:twoLinkArmCtrl2}
}
\caption{Two link arm tracking feedback control laws. In red and dashed the open loop
  (feedforward) law; in blue and solid, the feedback law.}
\label{fig:twoLinkArmControls}
\end{figure}

\begin{figure}[!h]
\subfloat[Tracking feedback error $h_x - h_{xr}$.]{%
  \includegraphics[width=.5\linewidth,clip,trim=20 180 40 220]{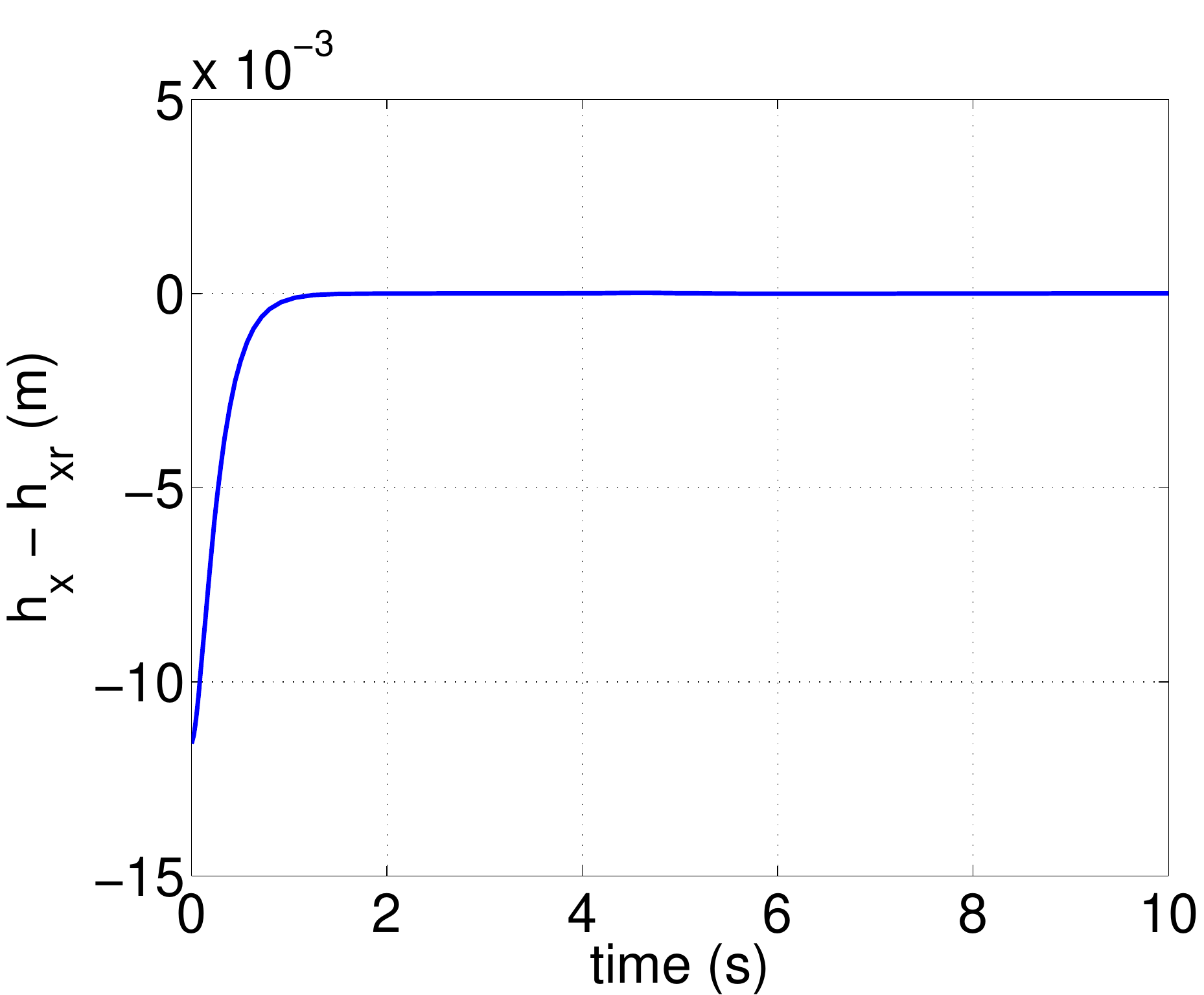}
  \label{fig:twoLinkArmErrHx} 
}
\subfloat[Tracking feedback error $h_y - h_{yr}$.]{%
  \includegraphics[width=.5\linewidth,clip,trim=20 180 40 220]{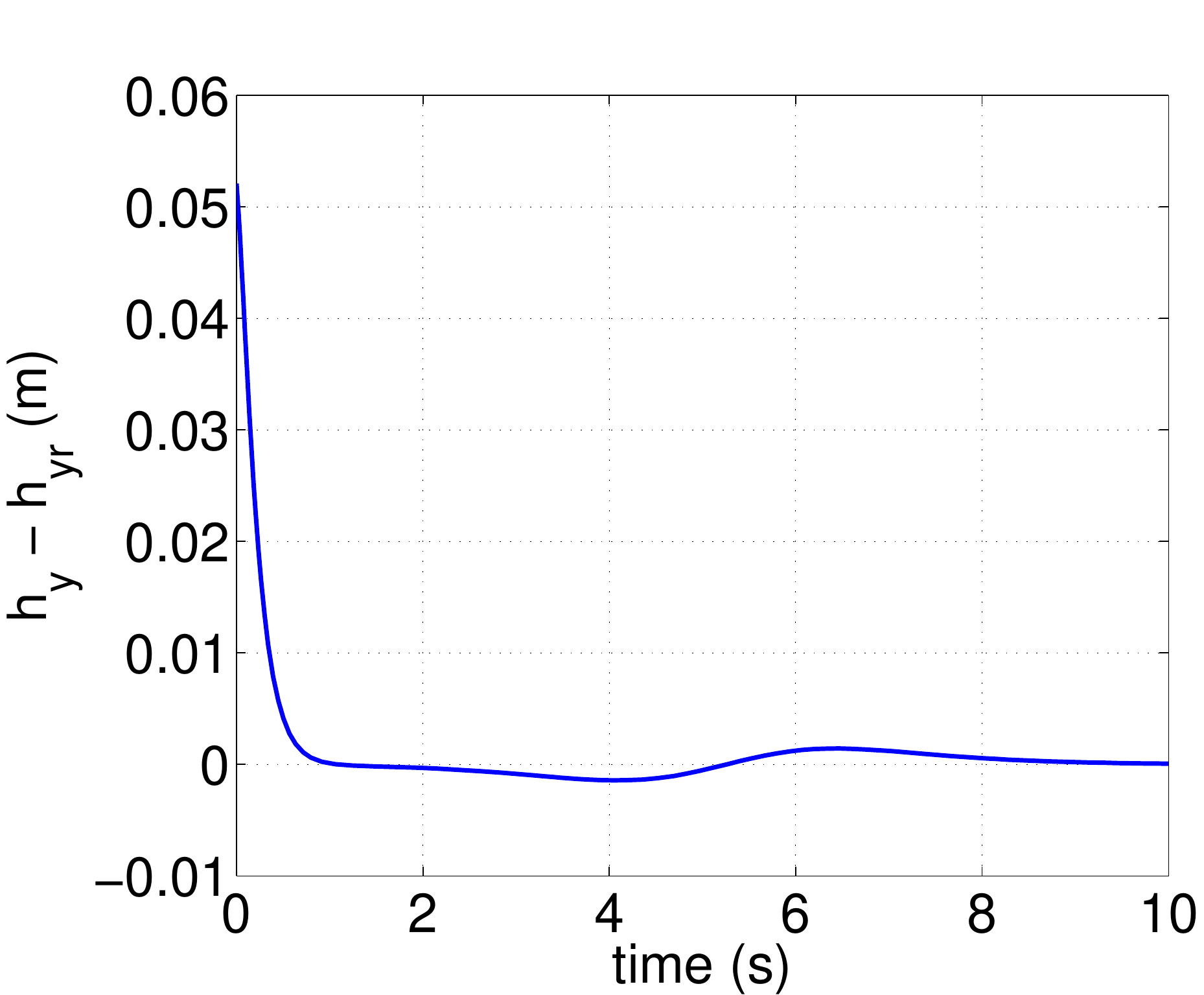}
  \label{fig:twoLinkArmErrHy}
}
\caption{Two link arm tracking feedback errors.}
\label{fig:twoLinkArmErrH}
\end{figure}

The trajectory tracking is illustrated in Figure \ref{fig:twoLinkArmHxHy}, and the 
associated animation in Figure \ref{fig:twoLinkArmAnimation}.
The corresponding control laws are shown in Figures \ref{fig:twoLinkArmCtrl1} and 
\ref{fig:twoLinkArmCtrl2}.
The tracking errors are plotted in Figures \ref{fig:twoLinkArmErrHx} and \ref{fig:twoLinkArmErrHy}.

 The previous results can be applied to more complete and less simplistic models, 
 as in e.g. \cite{friston-2010}; see also \cite{richardsonEtAl-MotorControl-2013}.

\section{Extensions of differential flatness} 
Several extensions of differentially flat systems can be envisioned. The recent paper
\cite{aschenbrennerEtAlNDJFL2013} reviews some of the most interesting ones. 
A Liouvillian closed structure $\mathfrak{D}$ will contain all the solutions 
of first order linear differential equations, an extended Liouvillian one will contain 
all the solutions of linear differential equations and an existentially closed one all the 
solutions of algebraic differential equations. We shall give some rather elementary definitions
below (see Appendix \ref{app:defFlatnessExensions}, p.~\pageref{app:defFlatnessExensions}).
\begin{definition}
The system
\begin{align*}
  \dot{\Vect{x}} &= f (\Vect{x},\Vect{u})
\end{align*}
with $\Vect{x}(t) \in \Reals^n$ and $\Vect{u}(t) \in \Reals^m$ is called
\emph{Liouvillian} (resp. \emph{extended Liouvillian, existentially closed}) if there exists a set 
of variables, called a \emph{Liouvillian} (resp. \emph{extended Liouvillian, existentially closed})
\emph{output} $\Vect{y} = (y_1, \ldots, y_m)$ solution of
\begin{align}
  \label{Eq:sortieLiouvillienne}
H(\Vect{y}, \dot{\Vect{y}}, \ldots, \Vect{y}^{(r_y)}, \Vect{x}, \Vect{u}, \dot{\Vect{u}}, \ldots,
\Vect{u}^{(r_u)}) &= 0, \qquad  r_y, r_u \in \Naturals
\end{align}
with $H$ linear of first order (resp. linear, polynomial) in its variables, such that
\begin{align*}
 \Vect{x} &= A (\Vect{y}, \dot{\Vect{y}}, \ldots, \Vect{y}^{(\rho_x)})\\
 \Vect{u} &= B (\Vect{y}, \dot{\Vect{y}}, \ldots, \Vect{y}^{(\rho_u)} )
\end{align*}
with $q$ an integer, and such that the system equations
\begin{align*}
\hspace*{-3ex}
&\dfrac{dA}{dt}(\Vect{y}, \dot{\Vect{y}}, \ldots, \Vect{y}^{(q+1)}) = \\
&\hspace*{9ex}
f(A(\Vect{y}, \dot{\Vect{y}}, \ldots, \Vect{y}^{(q)}), B(\Vect{y},
  \dot{\Vect{y}}, \ldots, \Vect{y}^{(q+1)}))
\end{align*}
are identically satisfied.
\end{definition}
\if@twocolumn
  \vfill\vfill
\else \fi

\section[Flatness of other neural mass models]{Differential flatness of some other neural mass models}
\subsection{Jansen and Rit model}
\subsubsection*{Brief recall of the model}
Consider the Jansen and Rit model, as depicted in \cite{pinotsisEtAl-2012}:
\begin{subequations}
\label{eq:jansenAndRitModel}
 \begin{align}
 \label{eq:jansenAndRitModelOne}
 \ddot \nu_1 + 2 \kappa_e \dot \nu_1 + \kappa_e^2 \nu_1 &= 
     \kappa_e m_e \big( w_{13} F(\nu_3) + u \big) \\
 \label{eq:jansenAndRitModelTwo}
 \ddot \nu_2 + 2 \kappa_i \dot \nu_2 + \kappa_i^2 \nu_2 &= 
     \kappa_i m_i w_{23} F(\nu_3) \\
 \label{eq:jansenAndRitModelThree}
 \ddot \nu_3 + 2 \kappa_e \dot \nu_3 + \kappa_e^2 \nu_3 &= 
     \kappa_e m_e \big( w_{31} F(\nu_1) + w_{32} F(\nu_2)  \big)  \\
 \label{eq:jansenAndRitModelFour}
 y &= \nu_3
 \end{align} 
\end{subequations}
These equations respectively depict the following populations:
excitatory stellate, inhibitory and excitatory. The signification of the various
variables and parameters are the following:
\begin{compactItemize}
\item[$\nu_i$] Expected depolarization in the $i$-th population 
\item[$w_{ij} F(\nu_j)$] presynaptic input to the $i$-th population from the $j^{\text{th}}$ one 
\item[$F(\nu_j)$] Sigmoid function of the postsynaptic depolarization 
\item[$w_{ij}$] Instrinsic connection strength between the populations $j$ and $j$ 
\item[$m_i, m_e$] Maximum postsynaptic responses 
\item[$\kappa_e, \kappa_i$] Rate constants of postsynaptic filtering 
\item[$u$] Exogenous input 
\item[$y$] Endogenous output 
\end{compactItemize}

\par\vspace*{1ex}
The choice made in \cite{pinotsisEtAl-2012} for the sigmoid function $F$ is the logistic
function
\begin{align*}
    F(\nu) &= \dfrac{1}{1 + e^{-\beta (\nu - \nu_T)}}
\end{align*}
whose derivative and inverse are:
\if@twocolumn
\begin{alignat*}{2}
   F' &= \beta F(F-1), &\quad \text{and} \quad F^{-1} (\eta) &= \phi (\eta) \\
   && &= \nu_T + \dfrac{1}{\beta}\, \text{ln} \dfrac{\eta}{\eta - 1}
\end{alignat*}
\else
\begin{align*}
   F' &= \beta F(F-1), \quad \text{and} \quad F^{-1} (\eta) = \phi (\eta) 
   = \nu_T + \dfrac{1}{\beta}\, \text{ln} \dfrac{\eta}{\eta - 1}
\end{align*}
\fi
Let $\mathsf{d}_i$, $\mathsf{d}_e$ be the differential operators
\begin{align*}
 \mathsf{d}_i &= \dfrac{d^2}{dt^2} + 2 \kappa_i \dfrac{d}{dt} + \kappa_i^2 = 
    \left( \dfrac{d}{dt} + \kappa_i \right)^2 \\
 \mathsf{d}_e &= \left( \dfrac{d}{dt} + \kappa_e \right)^2
\end{align*}
Then, the previous model \eqref{eq:jansenAndRitModel} can be written as
\begin{subequations}
\label{eq:jansenAndRitModelCompact}
 \begin{align}
 \label{eq:jansenAndRitModelCompactOne}
 \mathsf{d}_e \nu_1 &= 
     \kappa_e m_e \big( w_{13} F(\nu_3) + u \big) \\
 \label{eq:jansenAndRitModelCompactTwo}
 \mathsf{d}_i \nu_2 &= 
     \kappa_i m_i w_{23} F(\nu_3) \\
 \label{eq:jansenAndRitModelCompactThree}
 \mathsf{d}_e \nu_3 &= 
     \kappa_e m_e \big( w_{31} F(\nu_1) + w_{32} F(\nu_2)  \big)  \\
 \label{eq:jansenAndRitModelCompactFour}
 y &= \nu_3
 \end{align} 
\end{subequations}

\subsubsection*{Differential flatness of the model}
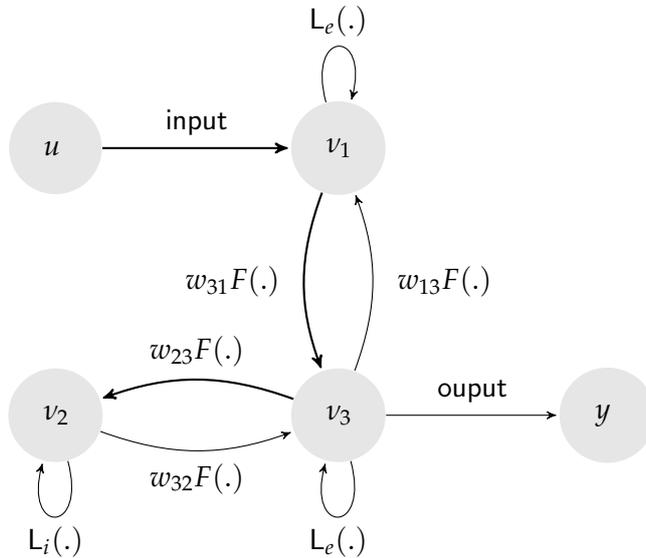
\begin{figure}[h!]
\begin{center}
\begin{tikzpicture}[->,>=stealth',shorten >=1pt,auto,node distance=3cm,%
main node/.style={circle,fill=black!10,font=\sffamily\normalsize\bfseries}]
  \node[main node] (1) {\begin{minipage}{5ex}\ \ $u$  \end{minipage}};
  \node[main node] (2) [right=4ex,right of=1] {%
                        \begin{minipage}{5ex}\ \ $\nu_1$ \end{minipage}};
  \node[main node] (3) [below=3ex,below of=2] {%
                        \begin{minipage}{5ex}\ \ $\nu_3$ \end{minipage}};
  \node[main node] (4) [below=3ex,below of=1] {%
                        \begin{minipage}{5ex}\ \ $\nu_2$  \end{minipage}};
  \node[main node] (5) [right=3ex,right of=3] {%
                        \begin{minipage}{5ex}\ \ $y$    \end{minipage}};
  \path[every node/.style={font=\sffamily\normalsize}]
    (1) edge  [thick]              node [above=0.2ex] {input}              (2)
    (2) edge [thick,bend right=20] node [right=6ex]   {$w_{13} F(.)$}      (3)
    (2) edge [loop above]          node               {$\mathsf{L}_e (.)$} (2)
    (3) edge [bend right=20]       node [left=6ex]    {$w_{31} F(.)$}      (2)
    (3) edge [thick,bend right=20] node [above=0.2ex] {$w_{23} F(.)$}      (4)
    (3) edge [loop below]          node               {$\mathsf{L}_e (.)$} (3)
    (4) edge [bend right=20]       node [below=0.2ex] {$w_{32} F(.)$}      (3)
    (4) edge [loop below]          node               {$\mathsf{L}_i (.)$} (4)
    (3) edge                       node [above=0.2ex] {ouput}              (5);
\end{tikzpicture}
\caption{\label{fig:jansenAndRitModel}The Jansen and Rit Model.}
\end{center}
\end{figure}

A flat output of the model \eqref{eq:jansenAndRitModel} is $\nu_2$. Indeed, 
after equation \eqref{eq:jansenAndRitModelTwo}, one gets $\nu_3$:
\begin{align}
 \label{eq:jansenAndRitModelFlatnessNu3}
 \nu_3 &= \phi \left( \dfrac{1}{\kappa_i m_i w_{23}}\, 
   (\ddot \nu_2 + 2 \kappa_i \dot \nu_2 + \kappa_i^2 \nu_2) \right)
\end{align}
Then, after \eqref{eq:jansenAndRitModelThree} 
\begin{align*}
 w_{31} F(\nu_1) &= w_{32} F(\nu_2) + \dfrac{1}{\kappa_e m_e}\, 
   (\ddot \nu_3 + 2 \kappa_e \dot \nu_3 + \kappa_e^2 \nu_3)
\end{align*}
Hence the expression for $\nu_1$:
\begin{align}
 \hspace*{-3ex}
 \label{eq:jansenAndRitModelFlatnessNu1}
 \nu_1 &= \phi \left[ \dfrac{w_{32}}{w_{31}}\, F(\nu_2) + \dfrac{1}{\kappa_e m_e w_{31}}\, 
   (\ddot \nu_3 + 2 \kappa_e \dot \nu_3 + \kappa_e^2 \nu_3) \right]
\end{align}
And, using \eqref{eq:jansenAndRitModelOne}, the expression for $u$:
\begin{align}
 \label{eq:jansenAndRitModelFlatnessU}
 u &= - w_{13} F (\nu_3) + \dfrac{1}{\kappa_e m_e} 
    (\ddot \nu_1 + 2 \kappa_e \dot \nu_1 + \kappa_e^2 \nu_1)
\end{align}
The Figure \ref{fig:jansenAndRitModel} below outlines the 
compartmental like model underlying the model \eqref{eq:jansenAndRitModel}.
The bold arrows in this Figure enables one, by reversing the arrows, to reveal 
the differential flatness character of the model: $\nu_3$ is obtained from
$\nu_2$ by reversing the arrow $(\nu_3) \rightarrow (\nu_2)$ (yielding equation
\eqref{eq:jansenAndRitModelFlatnessNu3}) ; then, $\nu_1$ is obtained from
$\nu_3$ (and $\nu_2$) by reversing the arrow $(\nu_1) \rightarrow (\nu_3)$ (yielding 
equation \eqref{eq:jansenAndRitModelFlatnessNu1}) ; finally $u$ is obtained from
$\nu_1$ (and $\nu_3$) by reversing the arrow $(u) \rightarrow (\nu_1)$ (yielding 
equation \eqref{eq:jansenAndRitModelFlatnessU}).

\subsubsection*{Extended Liouvillian character of the model}
Since the output of interest considered in \cite{pinotsisEtAl-2012} is $\nu_3$,
we can investigate how the model can be parametrized by this variable. The variable
$\nu_2$ can be obtained from $\nu_3$ by integrating the differential equation 
\eqref{eq:jansenAndRitModelTwo} in $\nu_2$ (which is linear in this variable).
\if@twocolumn
  \vfill\vfill
\else \fi
Indeed, \eqref{eq:jansenAndRitModelTwo} can be rewritten as:
\begin{align*}
 \dfrac{d}{dt}\, \begin{pmatrix}
                  \nu_2 \\ \dot \nu_2
                 \end{pmatrix} &=
    \begin{pmatrix}
     0 & 1 \\
     -\kappa_i^2 & -2 \kappa_i
    \end{pmatrix}
    \begin{pmatrix}
       \nu_2 \\ \dot \nu_2
    \end{pmatrix} +
    \begin{pmatrix}
      0 \\ \kappa_i m_i w_{23} F(\nu_3)
    \end{pmatrix}
\end{align*}
or, in matrix form
\begin{align*}
  \dot V &= A V + U, \qquad \text{with } \\
  V &= \begin{pmatrix}
       \nu_2 \\ \dot \nu_2
    \end{pmatrix},\ A = \begin{pmatrix}
     0 & 1 \\
     -\kappa_i^2 & -2 \kappa_i
    \end{pmatrix},\ U =     \begin{pmatrix}
      0 \\ \kappa_i m_i w_{23} F(\nu_3)
    \end{pmatrix}
\end{align*}
The general solution of this last equation is well known to be
\begin{align*}
 V &= V(0) e^{A t} + \int_0^t e^{A (t-\tau)} U (\tau) d\tau 
\end{align*}
One has
\begin{align*}
 e^{A t} &= \begin{pmatrix}
             (1 + \kappa_i t) e^{-\kappa_i t} &  t e^{-\kappa_i t} \\
             -\kappa_i^2 e^{-\kappa_i t} & (1 - \kappa_i t) e^{-\kappa_i t}
            \end{pmatrix}
\end{align*}
Thus, $\nu_2$ is given by
\begin{align}
\hspace*{-2ex}
\label{eq:jansenAndRitModelLiouvillainNu2}
  \nu_2 &= \nu_{20} \big( (1 + \kappa_i t ) e^{-\kappa_i t} \big) +
    \dot \nu_{20} t e^{-\kappa_i t} + \notag \\
    &\hspace*{4ex} \kappa_i m_i w_{23}
    \int_ 0^t (t-\tau ) e^{-\kappa_i (t-\tau )} F(\nu_3 (\tau)) d\tau
\end{align}
where $\nu_{20} = \nu_2 (0)$, $\dot \nu_{20} = \dot \nu_2 (0)$.
Then, the two other variables are obtained as in 
\eqref{eq:jansenAndRitModelFlatnessNu1}--\eqref{eq:jansenAndRitModelFlatnessU}:
\begin{subequations}
\label{eq:jansenAndRitModelLiouvillain}
\begin{align}
 \hspace*{-3ex}
 \label{eq:jansenAndRitModelLiouvillainNu1}
 \nu_1 &= \phi \left[ \dfrac{w_{32}}{w_{31}}\, F(\nu_2) + \dfrac{1}{\kappa_e m_e w_{31}}\, 
   (\ddot \nu_3 + 2 \kappa_e \dot \nu_3 + \kappa_e^2 \nu_3) \right] \\
 \label{eq:jansenAndRitModelLiouvillainU}
 u &= - w_{13} F (\nu_3) + \dfrac{1}{\kappa_e m_e} 
    (\ddot \nu_1 + 2 \kappa_e \dot \nu_1 + \kappa_e^2 \nu_1)
\end{align} 
\end{subequations}
Recalling $\mathsf{d}_i$, $\mathsf{d}_e$ the differential operators
\begin{align*}
 \mathsf{d}_i &= \left( \dfrac{d}{dt} + \kappa_i \right)^2 \qquad
 \mathsf{d}_e = \left( \dfrac{d}{dt} + \kappa_e \right)^2
\end{align*}
The previous equations 
\eqref{eq:jansenAndRitModelLiouvillainNu2}--\eqref{eq:jansenAndRitModelLiouvillain}
can be rewritten as:
\begin{subequations}\begin{align}
 \label{eq:jansenAndRitModelLiouvillainBisNu2}
 \nu_2 &= \mathsf{d}_i^{-1} \big( \kappa_i m_i w_{23} F(\nu_3) \big) \\
 \label{eq:jansenAndRitModelLiouvillainBisNu1}
 \nu_1 &= \phi \left[ \dfrac{w_{32}}{w_{31}}\, F(\nu_2) + \dfrac{1}{\kappa_e m_e w_{31}}\, 
   \mathsf{d}_e \nu_3 \right] \\
 \label{eq:jansenAndRitModelLiouvillainBisU}
 u &= - w_{13} F (\nu_3) + \dfrac{1}{\kappa_e m_e} \mathsf{d}_e \nu_1
\end{align}\end{subequations}
Thus, the model is extended Liouvillian and an extended Liouvillian output is $\nu_3$.

\cleardoublepage
\part{Neural field population models}

\subsection{General case model}
\label{sec:neuralFieldModels}
We consider spatially distributed network models, such as the ones
considered in Chapter 8 of \cite{ermentroutTerman-Springer-2010}, Subsection 8.4, p.~223, and
Chapter 12, Subsection 12.3.1, p.~376 or in Chapter 6 of \cite{bressloff-Springer-2014}, 
Subsection 6.5, p.~264 (as well as Subsection 2.5, p.~14 of \cite{bressloffJPhysA2012}).

We can consider the so-called activity-based neural field model:
\if@twocolumn
\begin{align*}
 \hspace*{-4ex}
 \tau_{sy} \dfrac{\partial \nu (t, x)}{\partial t}  &= - \nu (t, x) + 
         F\big( I_r w(x) \ast_x \nu (t,x) \big) h(\nu (t, x))
                 +\\
  &\hspace*{36ex} u(t, x)
\end{align*}
\else
\begin{align*}
 \tau_{sy} \dfrac{\partial \nu (t, x)}{\partial t}  &= - \nu (t, x) + 
         F\big( I_r w(x) \ast_x \nu (t,x) \big) h(\nu (t, x))
                 + u(t, x)
\end{align*}
\fi
or, in a slightly more compact way
\begin{align}
 \label{eq:generalPopulationDynamicsActivityModel}
 \tau_{sy} \dfrac{\partial \nu }{\partial t}  &= - \nu  + 
         F\big( I_r w(x) \ast_x \nu  \big) h(\nu )
                 + u
\end{align}
And a slightly more general case 
\begin{align}
 \hspace*{-4ex}
 \label{eq:moreGeneralPopulationDynamicsActivityModel}
 \tau_{sy} \dfrac{\partial \nu}{\partial t}  &= - \nu  + 
         F\left( I_r \int_{\Omega_x} w(x-\xi) \nu (t,\xi) d\xi \right) h(\nu )
                 + u
\end{align}
Alternately, we can consider the slightly different model
\begin{align}
 \label{eq:generalPopulationDynamicsModel}
 \tau_{sy} \dfrac{\partial \nu }{\partial t}  &= - \nu  + 
         I_r w(x) \ast_x F\big(\nu (t,x)\big) h(\nu )
                 + u
\end{align}
And its slight generalization 
\begin{align}
 \label{eq:moreGeneralPopulationDynamicsModel}
 \tau_{sy} \dfrac{\partial \nu }{\partial t}  &= - \nu  + 
          I_r \int_{\Omega_x} w(x-\xi) F\big(\nu (t,\xi)\big) d\xi h(\nu )
                 + u
\end{align}

\subsection{Parameters and variables assumptions}
The various parameters and functions staisfy the following:
\begin{itemize}[leftmargin=3ex]
 \item The spatial variable is three dimensional, i.e. $\Omega_x \subset \Reals^3$. 
 \item The parameter $\tau_{sy}$ is a constant. 
 \item If the synapses are saturating, then $h(s) = 1 - s$, otherwise, $h(s) = 1$.
 \item The function $F$ is a sigmoid type function (see \ref{subsec:sigmoidFunctions}, 
     p.~\pageref{subsec:sigmoidFunctions}).
 \item The control $u(t, r)$ has a spatial compact support $\Omega_u$.
 \item The neuron interaction strength function $w(r)$ is symmetric, nonegative, integrates to 1 
   over the whole line and is rapidly decaying at infinity:
\if@twocolumn
\begin{align*}
\hspace*{-5ex}
 \exists M &\in \Reals^3, \exists \alpha \in \Reals^+, \forall x \in \Reals^3 
     \text{ s.t. } \| x \| > M, \\
     &\hspace*{30ex}\| w(x) \| < \| x^{-\alpha} \|
\end{align*}
\else
\begin{align*}
 \exists M &\in \Reals^3, \exists \alpha \in \Reals^+, \forall x \in \Reals^3 
     \text{ s.t. } \| x \| > M, \| w(x) \| < \| x^{-\alpha} \|
\end{align*}
\fi
\end{itemize}
\subsection{Neuron interaction strength examples}
Some typical examples of such $w$ functions are shown in Table 
\ref{tab:NeuronInteractionFunctions}.
\begin{table}
\begin{center}
\begin{tabular}{c<{\hspace*{-1.5ex}}l<{\hspace*{-1.5ex}}l}
\toprule
Acronym & Name & Function $w$ \\
\midrule \par
(Wdorg) & \emph{Dirac at the origin} & $\delta_0$ \\[.4ex]
(Wdnor) &  \emph{Dirac not at the origin} & $\delta_{x_0}$ \\[.4ex]
(Wsofd) & \emph{Sum of Diracs} & $\sum_{i=1}^N a_i \delta_{x_i}$ \\[.4ex]
(Wsexp) &  \emph{Single exponential} & $e^{-ax} H(t)$ \\[.4ex]
(Wmexp) & \emph{Multiple exponential} & $\sum_{i=1}^N e^{-a_i x} H(t)$ \\[.4ex]
(Wgaus) & \emph{Gaussian} & $e^{-x^2 /\sigma^2}$ \\[.4ex]
(Waexp) & \emph{Absolute exponential} & $e^{-|x|/2}$ \\[.4ex]
(Wdosc) & \emph{Decaying oscillatory} & $e^{-b |x|} (b \sin |x| + \cos x)$ \\[.4ex]
(Wfhat) & \emph{Flat hat shaped} & 
           $\text{rect}(x/\chi)$ \\[.4ex]
\if@twocolumn
(Wmhat) & \emph{Mexican hat} & $\Gamma_1 e^{-\gamma_1 x} - \Gamma_2 e^{-\gamma_2 x}$ \\
        &&    or \\
        &&  $e^{-c_1 x^2 / \sigma_1^2} / \sigma_1^2 -$\\
        &&  \hspace*{7ex} $e^{- c_2 x^2 / \sigma_2^2} / \sigma_2^2$ 
            \\[.4ex]    
\else 
(Wmhat) & \emph{Mexican hat} & $\Gamma_1 e^{-\gamma_1 x} - \Gamma_2 e^{-\gamma_2 x}$ or \\
        &&  $e^{-c_1 x^2 / \sigma_1^2} / \sigma_1^2 - e^{- c_2 x^2 / \sigma_2^2} / \sigma_2^2$ 
            \\[.4ex]    
\fi
(Wwhat) & \emph{Wizard hat} & $(1/4) (1 - |x|) e^{-|x|}$ \\[.4ex]
(Wcomp) & \emph{Compact support} &  $w$ has compact support\\
\bottomrule
\end{tabular} 
\end{center}
\caption{Neuron interaction strength functions examples.}
\label{tab:NeuronInteractionFunctions}
\end{table} 

\begin{figure}[!h]
\subfloat[Absolute exponential.]{%
  \includegraphics[width=.5\linewidth,clip,trim=30 180 40 250]{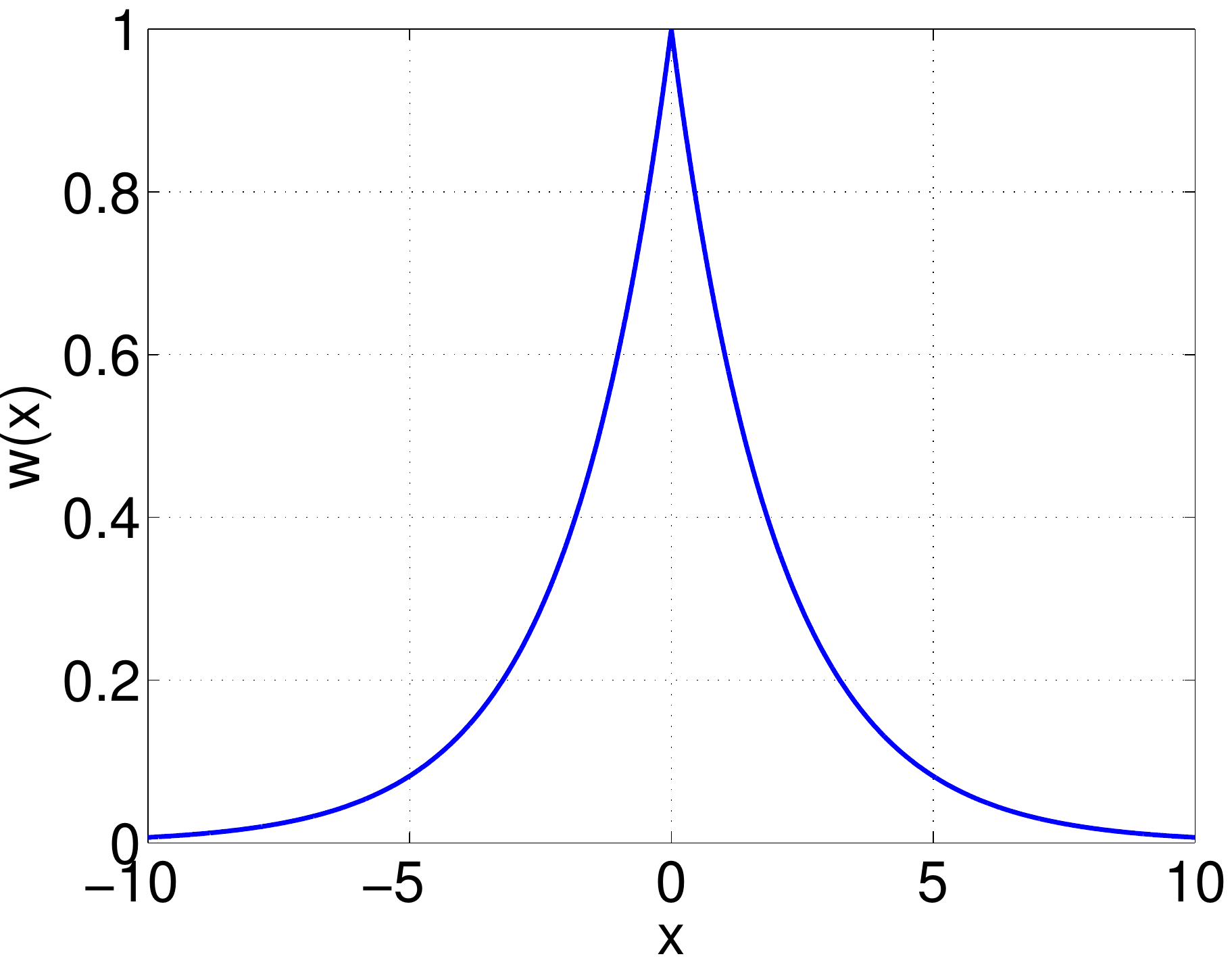}
  \label{fig:absExp} 
}
\subfloat[Decaying oscillatory with $b = 0.3$.]{%
  \includegraphics[width=.5\linewidth,clip,trim=30 180 40 250]{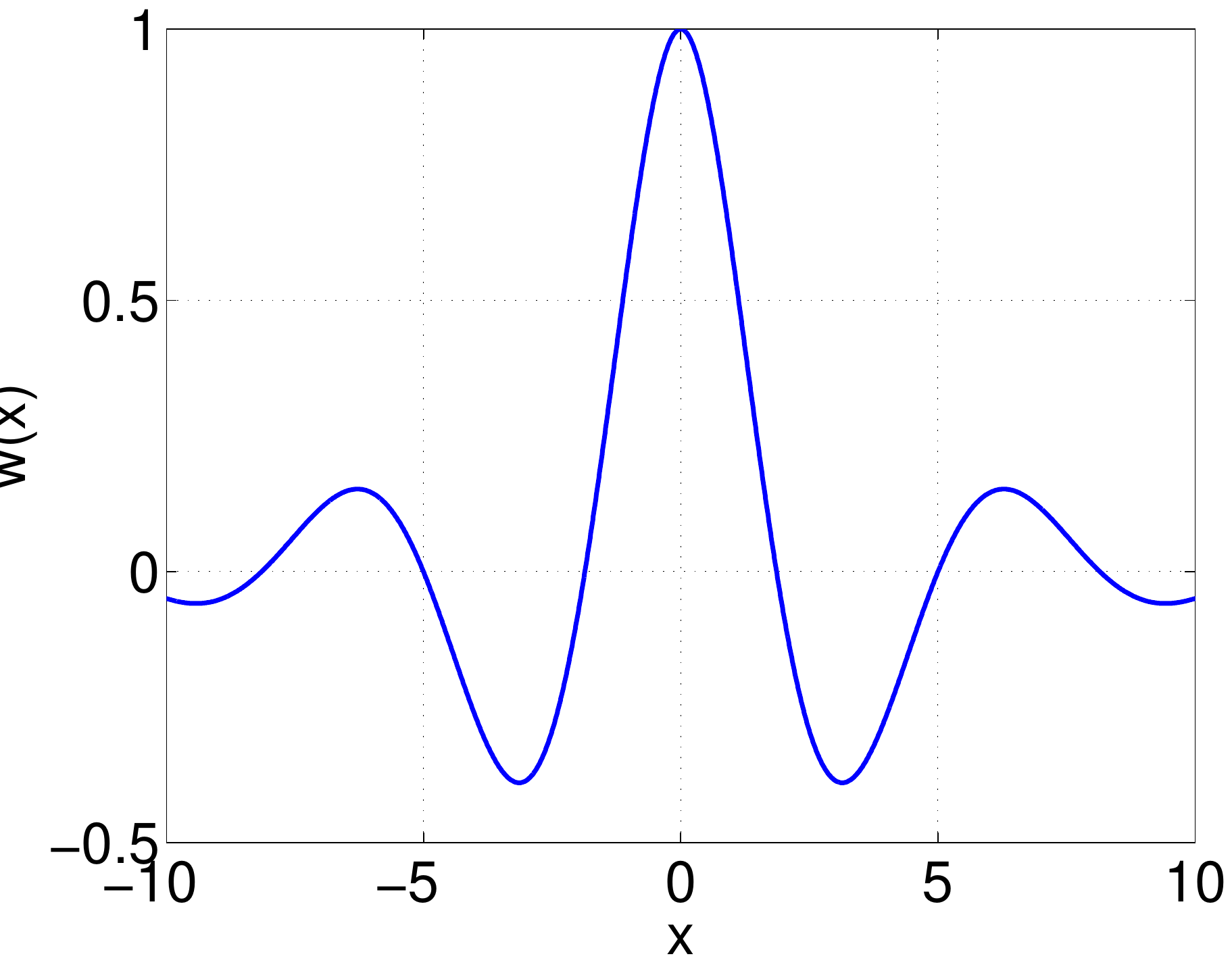}
  \label{fig:decayExp}
}
\caption{Examples of neuron interaction strength functions.}
\label{fig:NeuronInteractionFunctionsOne}
\end{figure}

\begin{figure}[!h]
\subfloat[Mexican hat through gaussian difference with $c_1 = 1, c_2 = 1.2, m_1 = 1, m_2 = 2$.]{%
  \includegraphics[width=.5\linewidth,clip,trim=30 180 40 250]{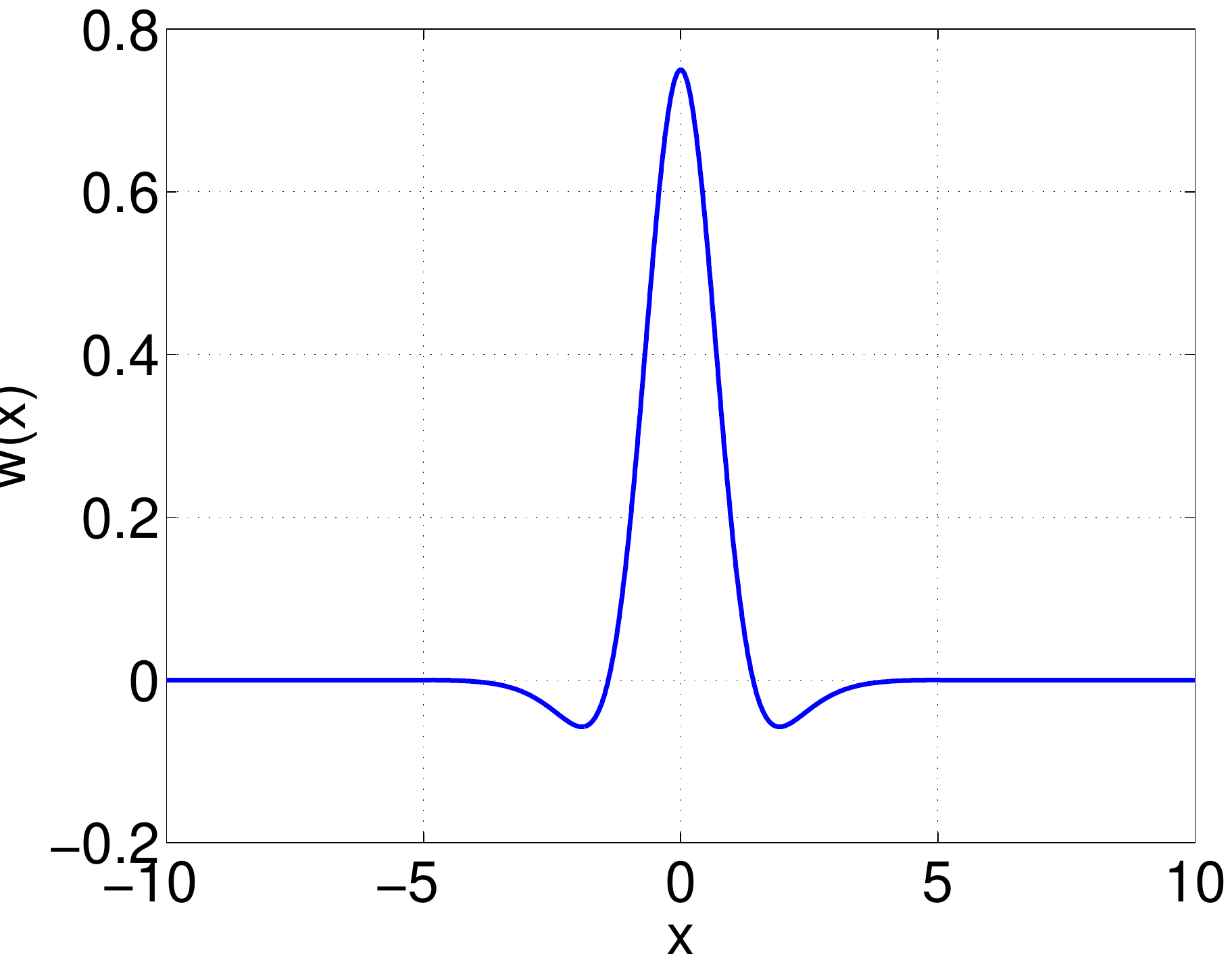}
  \label{fig:mexHat}
}
\subfloat[Wizard hat.]{%
\includegraphics[width=.5\linewidth,clip,trim=30 180 40 250]{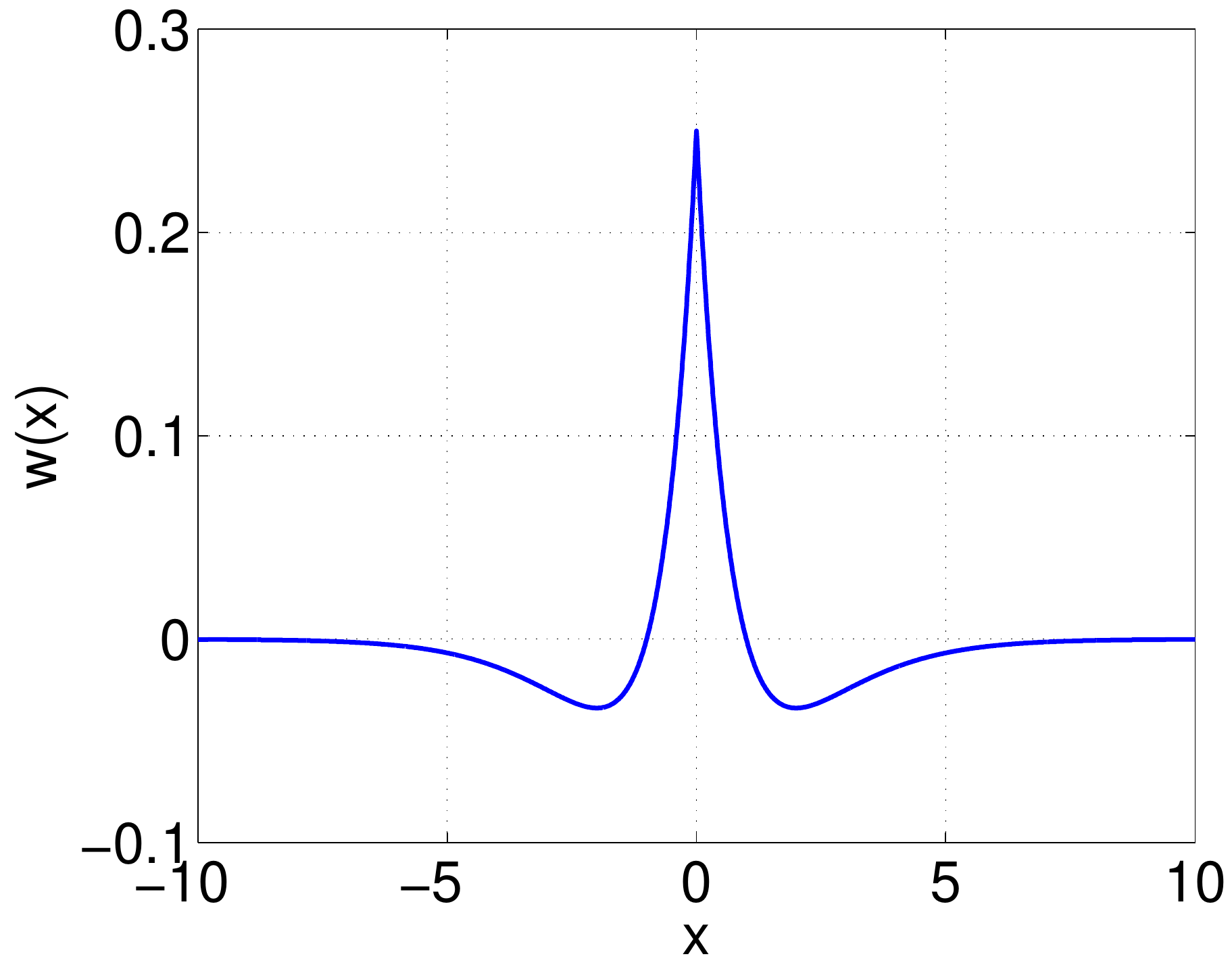}
  \label{fig:wizardHat}
}
\caption{Examples of neuron interaction strength functions.}
\label{fig:NeuronInteractionFunctionsTwo}
\end{figure}

%
The graphical representations of some of the above mentionned neuron interaction 
strength functions are given in Figures \ref{fig:NeuronInteractionFunctionsOne}
and \ref{fig:NeuronInteractionFunctionsTwo}.

\subsection{Simplification hypotheses}
We shall make the following simplifying assumptions: 
\begin{itemize}
 \item[(H1)] We consider a spherical symmetric case, which boils down to the unidimensional 
     case where $\Omega_x = [a, b]$ in $r$ and $r = \| x \|$.
 \item[(H2)] We consider $F$ equal to a heaviside:
   \begin{align*}
     \forall \xi &\in \Reals^-, \quad F(\xi) = 0, \qquad
     \forall \xi \in \Reals^+, \quad F(\xi) = 1
   \end{align*}
 \item[(H3)] We consider that the synapses are not saturating, hence $h(s) = 1$.
\end{itemize}

\subsection{Pointwise neuron interaction strength models}

\subsubsection*{Dirac at the origin case}
We consider that $w$ is a single spatial Dirac:
\begin{align*}
     w(r) &= \delta_{0}
\end{align*}
We then obtain the following system
\begin{align*}
 \dot \nu (t, r) &= -a \nu (t, r) + \delta_{0} \ast_r \big(\mathbb{1}_{[a,b]} \nu (t, r)\big) + u(t, r) 
\end{align*}
Or, in other words
\begin{align}
 \label{eq:simplisticPopulationDynamicsModel}
 \dot \nu (t, r) &= -a \nu (t, r) + \mathbb{1}_{[a,b]} (r)\, \nu (t, r) + u(t, r) 
\end{align}

\subsubsection*{Dirac not at the origin case}
We consider that $w$ is a single spatial Dirac:
\begin{align*}
     w(t, r) &= \delta_{r_0}
\end{align*}
   with $r_0 \in [a, b]$.
We then obtain the following system
\begin{align}
 \dot \nu (t, r) &= -a \nu (t, r) + \delta_{r_0} \ast_r \big(\mathbb{1}_{[a,b]} \nu (t, r)\big) + u(t, r) 
\end{align}
Or, in other words
\begin{align}
 \hspace*{-3ex}
 \label{eq:simplisticPopulationDynamicsModelDirac}
 \dot \nu (t, r) &= -a \nu (t, r) + \mathbb{1}_{[a,b]} (r-r_0)\, \nu (t, r-r_0) + u(t, r) 
\end{align}

\subsection{Exponential type neuron interaction strength}
\subsubsection*{A single exponential}
Consider that the neuron interaction strength $w$ satisfies a linear differential equation:
\begin{align*}
 w'(r) &= -a w(r)
\end{align*}
The model is the following
\begin{align}
 \label{eq:PopulationDynamicsModel}
 \tau_{sy} \dfrac{\partial \nu (t, r)}{\partial t}  &= - \nu (t, r) + 
         I_r w(r) \ast_r \nu (t,r) + u(t, r)
\end{align}
Hence, the convolution part is
\begin{align*}
 w(r) \ast_r \nu (t,r) &= \dfrac{1}{I_r}\, \left( \tau_{sy} \dfrac{\partial \nu (t, r)}{\partial t} +
     \nu (t, r) - u(t,r) \right)
\end{align*}
And is spatial derivative is
\begin{align*}
 \hspace*{-4ex}
 w'(r) \ast_r \nu (t,r) &= \dfrac{1}{I_r}\, \left( \tau_{sy} 
    \dfrac{\partial^2 \nu (t, r)}{\partial t \partial r} +
    \dfrac{\partial \nu (t, r)}{\partial r} + \dfrac{\partial u(t, r)}{\partial r} 
     \right) \\
 &= -a w(r) \ast_r \nu (t,r) \\
 &= \dfrac{-a}{I_r}\, \left( \tau_{sy} \dfrac{\partial \nu (t, r)}{\partial t} +
     \nu (t, r) - u(t,r) \right)
\end{align*}
Hence, $\nu$ satisfies the following partial differential equation:
\begin{align*}
 \hspace*{-4ex}
 \tau_{sy} \partial_t \partial_r \nu (t, r)
 &=  -(a \tau_{sy}\partial_t + \partial_r + a) \nu (t, r)\! -\! (\partial_r + a) u(t,r) 
\end{align*}

\subsubsection*{A more general case}
In \cite{coombesEtAl-Springer-2014}, Chapter 5, ``PDE Methods for Two-Dimensional
Neural Fields'' by Carlo R. Laing, the case of neuron interaction strength $w$ with
a rational Fourier transform is considered:
\begin{align*}
  \mathcal{F}(w) (\xi) &= \widetilde{w} (\xi) = 
  \int_{-\infty}^{+\infty} w(\tau) e^{-j \xi \tau} d\tau
  = \dfrac{p(\xi^2)}{q(\xi^2)}
\end{align*}
with $p$ and $q$ two polynomials. Then, equation \eqref{eq:generalPopulationDynamicsModel} with
$h(s) = 1$:
\begin{align*}
 \hspace*{-3ex}
 \tau_{sy} \dfrac{\partial \nu (t, x)}{\partial t}  &= - \nu (t, x) + 
         I_r w(x) \ast_x F\big(\nu (t,x)\big)
                 + u(t, x)
\end{align*}
becomes
\begin{align*}
 (\tau_{sy} \partial_t + 1) \tilde{\nu}(t, \xi)  &=  
         I_r \widetilde{w}(\xi)  \widetilde{F}(\nu)(t,\xi) + \tilde{u}(t, \xi) \\
      &= I_r \dfrac{p(\xi^2)}{q(\xi^2)}  \widetilde{F}(\nu)(t,\xi) + \tilde{u}(t, \xi)
\end{align*}
Thus, through multiplication by $q(\xi^2)$:
\begin{align*}
 \hspace*{-4ex}
 (\tau_{sy} \partial_t\! +\! 1)q(\xi^2) \tilde{\nu}(t, \xi)  &=  
         I_r p(\xi^2)  \widetilde{F}(\nu)(t,\xi)\! +\! q(\xi^2)\tilde{u}(t, \xi)
\end{align*}
which yields, in the spatial domain:
\begin{align*}
 \hspace*{-4ex}
 (\tau_{sy} \partial_t\! +\! 1)q(\partial_x^2) \nu (t, x)  &=  
         I_r p(\partial_x^2)  F(\nu)(t,x)\! +\! q(\partial_x^2)u(t, x)
\end{align*}
The Fourier tranforms of some of the neuron interaction strength functions
of Table \ref{tab:NeuronInteractionFunctions}, p.~\pageref{tab:NeuronInteractionFunctions}
are shown in Table \ref{tab:NeuronInteractionFunctionsFourierTransforms}.
%
\begin{table}
\begin{center}
\begin{small}
\begin{tabular}{>{\hspace*{-1ex}\centering}p{11ex}>{\hspace*{-2ex}\centering\arraybackslash$\displaystyle}p{13ex}<{$}>{\hspace*{-2ex}\centering\arraybackslash$\displaystyle}c<{$}}
\toprule
Name & \text{Function/}\text{Distribution} & \text{Fourier transform}\\
\toprule
\emph{Dirac at the origin}     & \delta_0                         & 1 \\ 
\specialrule{.05pt}{.4ex}{1ex}
\emph{Dirac not at the origin} & \delta_{x_0}                     & e^{-j \xi x_0}\\
\specialrule{.05pt}{.4ex}{1ex}
\emph{Sum of Diracs}           & \sum_{i=1}^N a_i \delta_{x_i}    & 
                                                             \sum_{i=1}^N a_i e^{-j \xi x_i} \\
\specialrule{.05pt}{.5ex}{.8ex}
\emph{Flat hat shaped}         & \text{rect}\left(\frac{x}{\chi}\right)  & \text{sinc}\\
\specialrule{.05pt}{.8ex}{.5ex}
\emph{Gaussian}                & e^{\frac{-x^2}{\sigma^2}}        & 
                                      \dfrac{\sigma}{\sqrt{2}}\, e^{-\frac{\sigma^2 \xi^2}{4}}\\
\specialrule{.05pt}{.5ex}{.5ex}
\emph{Absolute exponential}    & e^{-a|x|}                        & \frac{2a}{a^2 + \xi^2}\\
\specialrule{.05pt}{.5ex}{1ex}
\emph{Decaying oscillatory}    & e^{-b |x|} (b \sin |x| +$ $\ \ \cos x) & 
                                      \frac{4b(b^2 + 1)}{\xi^4 + 2(b^2 - 1)\xi^2 + (b^2 + 1)^2} \\
\specialrule{.05pt}{.5ex}{1ex}
\emph{Mexican hat}             & \Gamma_1 e^{-\gamma_1 |x|} -$ $ 
                               \Gamma_2 e^{-\gamma_2 |x|}                & 
                               2\, \frac{\Gamma_1 \gamma_1 (\gamma_2 + \xi^2) - 
                                 \Gamma_2 \gamma_2 (\gamma_1^2 + \xi^2)}{(\gamma_1^2 + \xi^2)(
                                 \gamma_2^2 + \xi^2)}\\
\specialrule{.05pt}{.5ex}{.5ex}
\emph{Wizard hat}              &  \frac{(1 - |x|) e^{-|x|}}{4}     & \frac{\xi^2}{(1+\xi^2)^2}\\[2ex]
\bottomrule
\end{tabular}  
\end{small}
\end{center}
\caption{Fourier tranforms of some neuron interaction strength functions.}
\label{tab:NeuronInteractionFunctionsFourierTransforms}
\end{table}

\section{A Jansen and Rit Neural field model}
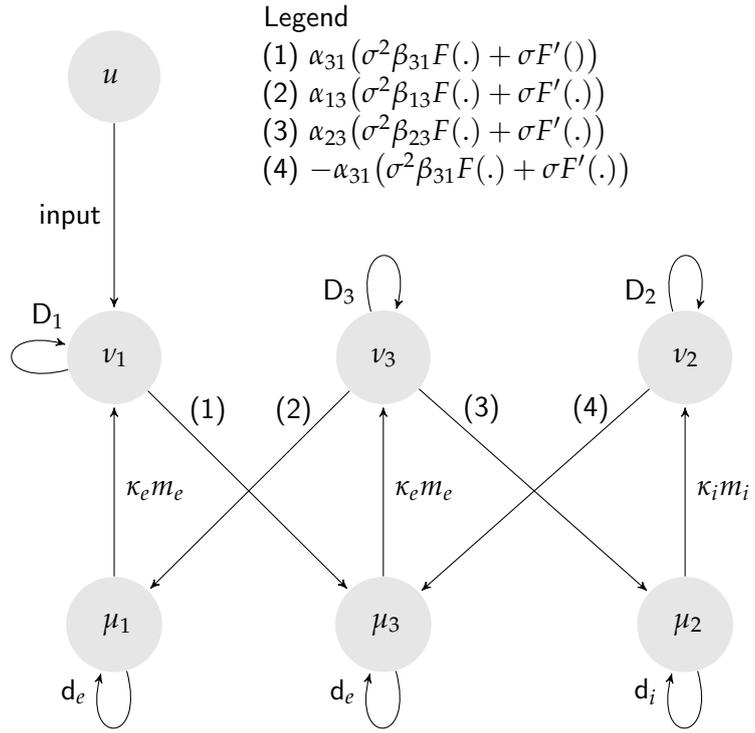
\begin{figure}[h!]
\begin{center}
\begin{tikzpicture}[->,>=stealth',shorten >=1pt,auto,node distance=3cm,%
main node/.style={circle,fill=black!10,font=\sffamily\normalsize\bfseries}]
  \node[main node] (u) {\begin{minipage}{5ex}\ \ $u$  \end{minipage}};
  \node[main node] (legend) [right=1ex,right of=u,fill=white] {};
  \node[main node] (nu1) [below=4ex,below of=u] {%
                        \begin{minipage}{5ex}\ \ $\nu_1$ \end{minipage}};
  \node[main node] (nu2) [right=25ex,right of=nu1] {%
                        \begin{minipage}{5ex}\ \ $\nu_2$ \end{minipage}};
  \node[main node] (nu3) [right=3ex,right of=nu1] {%
                        \begin{minipage}{5ex}\ \ $\nu_3$ \end{minipage}};
  \node[main node] (mu1) [below=3ex,below of=nu1] {%
                        \begin{minipage}{5ex}\ \ $\mu_1$ \end{minipage}};
  \node[main node] (mu2) [below=3ex,below of=nu2] {%
                        \begin{minipage}{5ex}\ \ $\mu_2$ \end{minipage}};
  \node[main node] (mu3) [below=3ex,below of=nu3] {%
                        \begin{minipage}{5ex}\ \ $\mu_3$ \end{minipage}};
  \path[every node/.style={font=\sffamily\normalsize}]
    (u)   edge  []                 node [left=0.2ex]  {input}                 (nu1)
    (legend)   []                  node [below=-6ex]  {%
       \hspace*{20ex}
       \begin{minipage}{35ex}
       Legend \\
       (1) $\alpha_{31} \big( \sigma^2 \beta_{31} F(.) + \sigma F'() \big)$  \\
       (2) $\alpha_{13} \big( \sigma^2 \beta_{13} F(.) + \sigma F'(.) \big)$ \\
       (3) $\alpha_{23} \big( \sigma^2 \beta_{23} F(.) + \sigma F'(.) \big)$ \\
       (4) $-\alpha_{31} \big( \sigma^2 \beta_{31} F(.) + \sigma F'(.) \big)$
       \end{minipage}}                                                          (u)
    (mu1) edge []             node [right=0.1ex]  {$\kappa_e m_e$}              (nu1)
    (mu2) edge []             node [right=0.1ex]  {$\kappa_i m_i$}              (nu2)
    (mu3) edge []             node [right=0.1ex]  {$\kappa_e m_e$}              (nu3)
    (mu1) edge [loop below]   node [pos=0.9,left=0.1ex]   {$\mathsf{d}_e$}      (mu1)
    (mu2) edge [loop below]   node [pos=0.9,left=0.1ex]   {$\mathsf{d}_i$}      (mu2)
    (mu3) edge [loop below]   node [pos=0.9,left=0.1ex]   {$\mathsf{d}_e$}      (mu3)
    (nu3) edge []             node [pos=0.1,left=0.5ex]   {(2)}                 (mu1)
    (nu3) edge []             node [pos=0.1,right=0.8ex]  {(3)}                 (mu2)
    (nu1) edge []             node [pos=0.1,right=0.8ex]  {(1)}                 (mu3)
    (nu2) edge []             node [pos=0.1,left=0.5ex]   {(4)}                 (mu3)
    (nu1) edge [loop left]    node [pos=0.9,above=0.2ex]   {$\mathsf{D}_1$}      (mu1)
    (nu2) edge [loop above]   node [pos=0.1,left=0.1ex]   {$\mathsf{D}_2$}      (mu2)
    (nu3) edge [loop above]   node [pos=0.1,left=0.1ex]   {$\mathsf{D}_3$}      (mu3);
\end{tikzpicture}
\caption{\label{fig:jansenAndRitNeuralFieldModel}A Jansen and Rit neural field model.}
\end{center}
\end{figure}

Let us consider, after \cite{pinotsisEtAl-2012}, the following neural field
Jansen and Rit model
\begin{subequations}
\label{eq:jansenAndRitNeuralFieldModel}
 \begin{align}
 \hspace*{-5ex}
 \label{eq:jansenAndRitNeuralFieldModelOne}
 \ddot \nu_1 + 2 \kappa_e \dot \nu_1 + \kappa_e^2 \nu_1 &= 
     \kappa_e m_e \big( \mu_1 + u \big) \\
 \label{eq:jansenAndRitNeuralFieldModelTwo}
 \ddot \nu_2 + 2 \kappa_i \dot \nu_2 + \kappa_i^2 \nu_2 &= 
     \kappa_i m_i \mu_2 \\
 \label{eq:jansenAndRitNeuralFieldModelThree}
 \ddot \nu_3 + 2 \kappa_e \dot \nu_3 + \kappa_e^2 \nu_3 &= 
     \kappa_e m_e \mu_3 \\
 \label{eq:jansenAndRitNeuralFieldModelFour}
 \partial_t^2 \mu_1 - \sigma^2 \partial_x^2 \mu_1 + 2 \sigma \beta_{13} \partial_t \mu_1 +
     \sigma^2 \beta_{13}^2 \mu_1 &= \phi_{13} (\nu_3) \\
 \label{eq:jansenAndRitNeuralFieldModelFive}
 \partial_t^2 \mu_2 - \sigma^2 \partial_x^2 \mu_2 + 2 \sigma \beta_{23} \partial_t \mu_2 +
     \sigma^2 \beta_{23}^2 \mu_2 &= \phi_{23} (\nu_3) \\
 \label{eq:jansenAndRitNeuralFieldModelSix}
 \partial_t^2 \mu_3 - \sigma^2 \partial_x^2 \mu_3 + 2 \sigma \beta_{31} \partial_t \mu_3 +
     \sigma^2 \beta_{31}^2 \mu_3 &= \psi_{31} (\nu_1, \nu_2)
 \end{align} 
\end{subequations}
with
\begin{align*}
  \phi_{ij} (\nu_j) &= \alpha_{ij} \left( \sigma^2 F(\nu_j) + \sigma F' (\nu_j) \right) \\
  \hspace*{-5ex}
  \psi_{31} (\nu_1, \nu_2) &= \alpha_{31} \Big( \sigma^2 \beta_{31} \big( F(\nu_1) - F(\nu_2) \big) + 
  \sigma \big( F'(\nu_1) - F'(\nu_2) \big) \Big)
\end{align*}
Let us recall the differential operators $\mathsf{d}_i$ and $\mathsf{d}_e$ and introduce
$\mathsf{D}_1$, $\mathsf{D}_2$, $\mathsf{D}_3$:
\begin{align*}
 \mathsf{d}_i &= \left( \dfrac{d}{dt} + \kappa_i \right)^2 \quad
       \mathsf{d}_e = \left( \dfrac{d}{dt} + \kappa_e \right)^2 
\end{align*}
\begin{align*}
 \mathsf{D}_1 &= \partial_t^2 - \sigma^2 \partial_x^2 + 2 \sigma \beta_{13} \partial_t  +
                 \sigma^2 \beta_{13}^2 \\
 \mathsf{D}_2 &= \partial_t^2 - \sigma^2 \partial_x^2 + 2 \sigma \beta_{23} \partial_t  +
                 \sigma^2 \beta_{23}^2, \\  
 \mathsf{D}_3 &= \partial_t^2 - \sigma^2 \partial_x^2 + 2 \sigma \beta_{31} \partial_t  +
                 \sigma^2 \beta_{31}^2                 
\end{align*}
The model \eqref{eq:jansenAndRitNeuralFieldModel} can then be rewritten as
\begin{subequations}
\label{eq:jansenAndRitNeuralFieldModelBis}
 \begin{align}
 \label{eq:jansenAndRitNeuralFieldModelBisOne}
 \mathsf{d}_e \nu_1 &= \kappa_e m_e \big( \mu_1 + u \big) \\
 \label{eq:jansenAndRitNeuralFieldModelBisTwo}
 \mathsf{d}_i \nu_2  &= \kappa_i m_i \mu_2 \\
 \label{eq:jansenAndRitNeuralFieldModelBisThree}
 \mathsf{d}_e \nu_3 &= \kappa_e m_e \mu_3 \\
 \label{eq:jansenAndRitNeuralFieldModelBisFour}
 \mathsf{D}_1 \mu_1  &= \phi_{13} (\nu_3) \\
 \label{eq:jansenAndRitNeuralFieldModelBisFive}
 \mathsf{D}_2 \mu_2  &= \phi_{23} (\nu_3)\\
 \label{eq:jansenAndRitNeuralFieldModelBisSix}
 \mathsf{D}_3 \mu_3  &= \psi_{31} (\nu_1, \nu_2)
 \end{align} 
\end{subequations}
The Figure \ref{fig:jansenAndRitNeuralFieldModel} below outlines the 
compartmental like model underlying the model \eqref{eq:jansenAndRitNeuralFieldModelBis}.

The variable $\nu_2$ is a Liouvillian output. Indeed, $\mu_2$ is obtained through $\nu_2$ 
using equation \eqref{eq:jansenAndRitNeuralFieldModelBisTwo}:
\begin{align}
 \mu_2 &= \dfrac{1}{\kappa_i m_i}\, \mathsf{d}_i \nu_2 
\end{align}
Then $\nu_3$ is obtained via $\mu_2$ with the help of 
\eqref{eq:jansenAndRitNeuralFieldModelBisFive}
\begin{align}
 \nu_3 &= \psi (\mathsf{D}_2 \mu_2)
\end{align}
where $\psi_2$ denotes the inverse function of $\alpha_{23} (\sigma^2 \beta_{23} F +
\sigma F' )$. After, $\mu_3$ is derived from $\nu_3$ through 
\eqref{eq:jansenAndRitNeuralFieldModelBisThree}
\begin{align}
 \mu_3 &= \dfrac{1}{\kappa_e m_e}\, \mathsf{d}_e \nu_3 
\end{align}
The variable $\mu_3$ and $\nu_2$ yields $\nu_1$ through \eqref{eq:jansenAndRitNeuralFieldModelBisSix}
\begin{align}
 \nu_1 &= \mathsf{D}_3 \mu_3 + \alpha_{31} \Big( \sigma^2 \beta_{31} 
     F(\nu_2)  + \sigma F'(\nu_2)  \Big) 
\end{align}
where $\psi_3$ is the inverse function of 
$\alpha_{31} ( \sigma^2 \beta_{31} F  + \sigma F' )$.
By integrating the wave equation in the equation \eqref{eq:jansenAndRitNeuralFieldModelBisFour}
we get $\mu_1$ from $\nu_3$
\begin{align}
 \mu_1 &= \alpha_{13}\, \mathsf{D}_1^{-1} \big( \sigma^2 \beta_{13} F(\nu_3) +
     \sigma F'(\nu_3) \big)
\end{align}
At last, the control input $u$ can de derived through \eqref{eq:jansenAndRitNeuralFieldModelBisOne}
from $\mu_1$ and $\nu_1$
\begin{align}
 u &= \dfrac{1}{\kappa_e m_e}\, \mathsf{d}_e \nu_1 - \mu_1 
\end{align}

%

\refstepcounter{dummy}
\addtocontents{toc}{\protect\vspace{\beforebibskip}} 
\addcontentsline{toc}{chapter}{\tocEntry{\bibname}}
\label{app:bibliography}

%

\cleardoublepage\part{Appendix}
\begin{appendix}
\chapter{Notations, transforms and sigmoid functions}
\label{app:notations}
\section{Functions and distributions}
\begin{itemize}
  \item The distribution $H(\eta)$ is the \emph{Heaviside} distribution:
   \begin{align*}
      H(\eta) &= \begin{cases}
                  0 \quad\text{ if } \eta \leqslant 0 \\
                  1 \quad\text{ if } \eta > 0               
              \end{cases}
   \end{align*}
  \item The function $\text{sinc}(\eta)$ is the \emph{cardinal sine}:
   \begin{align*}
      \text{sinc} (\eta) &= \dfrac{\sin (\eta)}{\eta}
   \end{align*}
  \item The distribution $\text{rect}(\eta)$ is the \emph{rectangular pulse} of width $1$:
   \begin{align*}
      \text{rect}(\eta) = \begin{cases}
              0 \quad \text{ if } |\eta| \geqslant \dfrac{1}{2} \\[1.5ex]
              1 \quad \text{ if } |\eta| < \dfrac{1}{2} 
              \end{cases}
   \end{align*}
  \item The \emph{boxcar} distribution (rectangular pulse of width $\rho$ and centered on $\eta_0$):
   \begin{align*}
      \text{rect}\left(\dfrac{t-t_0}{\rho} \right) = H(\eta - \eta_0 + \dfrac{\rho}{2}) -
         H(t - t_0 - \dfrac{\rho}{2}) 
   \end{align*}  
   \item  The \emph{linear rectifier} function is
   \begin{align*}
          |\eta|_+ &= \begin{cases}
                        \ 0 \quad &\text{when } v \leqslant 0 \\
                        \ v \quad &\text{when } v > 0  
                      \end{cases}
   \end{align*}
\end{itemize}
\section{Transforms}
Consider a function
$f(t, x)$ from $\Reals \times \Omega_x$ to $\Reals$, where $\Omega_x \subseteq \Reals^3$.
\begin{itemize}
 \item The function $\tilde{f} (t, \xi)$ will designate the \emph{spatial Fourier transform}
   of $f$, i.e.
   \begin{align*}
      \tilde{f} (t, \xi) &= \mathcal{F} (f) (t, \xi) = \int_{-\infty}^{+\infty} 
        f(t, x) e^{-j\xi x} dx
   \end{align*}
 \item The function $\hat{f} (s, x)$ will designate the \emph{temporal Laplace transform}
   of $f$, i.e.
   \begin{align*}
      \hat{f} (s, x) &= \mathcal{L} (f) (s, x) = \int_{-\infty}^{+\infty} 
        f(t, x) e^{-s t} dt
   \end{align*}
\end{itemize}
\section{Sigmoid functions}
\label{subsec:sigmoidFunctions}
The following functions $F$ will be in particular used for the firing rate. 
Thus $F(\xi)$ designates a spike rate and $\xi$ a sitmulus intensity.
A sigmoid function $F: \Reals \rightarrow \Reals$ is such that
\begin{align*}
 F(0) = F'(0) &= 0, \qquad F(1) = 1, \quad F'(1) = 0 \\
 \forall \xi \in \Reals, \xi < 0 \quad F(\xi) &= 0, \\
     \forall \xi \in \Reals, \xi > 1 \quad F(\xi) &= 1
\end{align*}
The following lists some of the most used sigmoid functions (see, e.g. 
\cite{ermentroutTerman-Springer-2010}, Section 11.3, p.~345; \cite{bressloff-Springer-2014},
pp.~9, 22, 254, 373, \cite{haken-Springer-2008}, pp.~ 14, 252).
\begin{itemize}
\item The \emph{Heaviside}.
  \begin{align*}
    F(x) &= F_0 H(\xi - \xi_0)
  \end{align*}
\item The \emph{Piecewise linear} function.
  \begin{align*}
    F(x) &= \begin{cases}
              \ 0 \quad &\text{if } \xi < \xi_0 \\
              \ \beta (\xi - \xi_0) \quad &\text{if } \xi_0 \leqslant \xi < \xi_0 + 1/\beta \\ 
              \ 0 \quad &\text{if } \xi > \xi_0 + 1/\beta
            \end{cases}
  \end{align*}
\item The \emph{Logistic} function.
  \begin{align*}
    F(x) &= \dfrac{1}{1 + e^{-\beta (x - x_T)}}
  \end{align*}
  one has
  \begin{align*}
   F' &= \beta F(F-1) \\
   F^{-1} (\eta) &= \phi (\eta) =
     x_T + \dfrac{1}{\beta}\, \text{ln} \dfrac{\eta}{\eta - 1}
  \end{align*}
\item The \emph{Traub Model}.
  \begin{align*}
    F(\xi) &= \dfrac{1}{1 + e^{\frac{-\xi-\beta}{\alpha}}}
  \end{align*}
\item The \emph{Hyperbolic tangent} function.
  \begin{align*}
   F(\xi) &= F_0 \big( 1 + \tanh (\alpha \xi) \big)
  \end{align*}
\item The \emph{Square root} function.
  \begin{align*}
   F(\xi) &= F_ 0 \sqrt{\xi -\xi_T}
  \end{align*}
\item The \emph{Noisy firing rate} function.
  \begin{align*}
    F(\xi) &= \sqrt{\dfrac{\xi - \xi_T}{1 - e^{\frac{-(\xi - \xi_T)}{\beta}}}}
  \end{align*}
  Here, $\beta$ is a measure of the noise, and when $\beta$ tends to zero, the 
  function approaches a pure square root model.
\item The \emph{Mean firing rate with flexible shape} function 
  (see \cite{coombesEtAl-Springer-2014}, p.~371).
  \begin{align*}
    F(\xi) &= F_m - \dfrac{F_m}{\big(1 + e^{\sqrt{2}\, \frac{\xi - \mu}{\sigma}}\big)^{\kappa}}
  \end{align*}
\item The \emph{Naka-Rushton} functions (alternately called Hill functions).
  \begin{align*}
    F(\xi) &= \begin{cases}
               \ \dfrac{r \xi^n}{\xi^n + \theta^n} \quad &\text{if } \xi \leqslant 0 \\
               \ 0  \quad &\text{if } \xi > 0
              \end{cases}    
  \end{align*}
  where $r$ is the maximum spike rate and $\theta$ is the value of the stimulus intensity
  for which $F$ reaches half its maximum. The exponent $n$ is a measure of the steepness of
  the $F(\xi)$ curve. Typical values matching experimental data range from $1.4$ to $3.4$.
  The function
  \begin{align*}
    F(\xi) &= 1 - \dfrac{\xi^n}{\xi^n + \theta} = \dfrac{\theta^n}{\xi^n + \theta^n}
  \end{align*}
  is also used.
\item The \emph{Algebraic sigmoid function}.
  \begin{align*}
    F(\xi) &= \dfrac{\xi}{\sqrt{1 + \xi^2}} 
  \end{align*}
  This function has the inverse
  \begin{align*}
    F^{1}(\eta) &= \dfrac{\eta}{\sqrt{1 - \eta^2}}
  \end{align*}
\end{itemize}

\chapter{Some flatness simple criteria}
There does not exist, at the time of this writing,
a general criterion for checking flatness, neither
for building flat outputs in a constructive manner.
Nevertheless, some peculiar cases are to be noticed.

\section{Necessary and sufficient conditions in peculiar cases}

\begin{proposition}
Any static state feedback linearizable system is flat.
\end{proposition}
See below (Subsection \ref{subsecStaticFeedbackLinearization}, 
p.~\pageref{subsecStaticFeedbackLinearization})
for a static state feedback linearizability
criterion for affine input systems.

\begin{proposition}[Charlet, Levine and Marino, 1989]
For systems with a single input, dynamic feedback linearization 
implies static feedback linearization.
\end{proposition}

\begin{proposition}[Charlet, Levine and Marino, 1989]
A dynamics affine in the input with $n$ states and $n-1$ inputs 
is flat as soon as it is controllable (strongly accessible).
\end{proposition}
Recall that a dynamics is called affine in the input if it is of
the form
\begin{align*}
\dot{\Vect{x}} &= f_0 (\Vect{x}) + \sum_{i=1}^{n-1} g_i (\Vect{x}) 
  \Vect{u}_i
\end{align*}
A dynamics with $\Vect{x} \in {\mathcal X} \subseteq \Reals^n$ is
strongly accessible if, for all $x \in {\mathcal X}$, there exists
a $T > 0$ such that
\begin{align*}
  \text{int}{\mathsf R}(\Vect{x}) \neq \varnothing
\end{align*}
where $\text{int}{\mathsf S}$ denotes the interior of the set 
${\mathsf S}$ and ${\mathsf R}(\Vect{x})$ is the reachable set
of $\Vect{x}$.

\section{A necessary condition}

\begin{proposition}[Ruled variety criterion, Rouchon, 1995]
Suppose the dynamics $\dot{\Vect{x}} = f(\Vect{x}, \Vect{u})$ is
flat. The projection of the sub variety $\Vect{p} = f(\Vect{x}, \Vect{u})$ 
in the $(\Vect{p}, \Vect{u})$-space ($\Vect{x}$ is here 
a parameter) onto the $\Vect{p}$-space is a ruled variety for
all $\Vect{x}$.
\end{proposition}
This criterion means that the elimination of $\Vect{u}$ from the
$n$ equations $\dot{\Vect{x}} = f(\Vect{x}, \Vect{u})$ yields 
$n-m$ equations $F(\Vect{x}, \dot{\Vect{x}}) = 0$ with the following
property: for all $(\Vect{x}, \Vect{p})$ such that $F(\Vect{x}, 
\Vect{p}) = 0$, there exists $\Vect{a} \in \Reals^n$, $\Vect{a} \neq 0$ 
such that
\begin{align*}
  \forall \lambda \in \Reals, \quad F(\Vect{x}, \Vect{p} + \lambda
    \Vect{a}) = 0
\end{align*}
The variety $F(\Vect{x}, {\Vect{p}}) $ is thus ruled since it
contains the line passing through $\Vect{p}$ with direction 
$\Vect{a}$.

\section{Static state feedback linearizability criterion}
\label{subsecStaticFeedbackLinearization}
Consider an affine input system 
\begin{align*}
\dot{\Vect{x}} &= f (\Vect{x}) + 
    \sum_{i=1}^{n-1} g_i (\Vect{x}) {u}_i =
    f (\Vect{x}) + g (\Vect{x}) \Vect{u}
\end{align*}
where $f, g_i$ are smooth vector fields on a domain $D \subset \Reals^n$, 
$\Vect{x} \in D$, $\Vect{u} \in \Reals^m$.

%

\subsection{Brief recall of differential geometry
notions}
\begin{definition}
  Let $r \geqslant 0$ be an integer. A \emph{$C^r$ vector field} on $\Reals^n$ is a
  mapping $f : D \rightarrow \Reals^n$ of class $C^r$ from an open set 
  $D \subset \Reals^n$ to $\Reals^n$. A \emph{smooth vector field} is a mapping 
  $f : D \rightarrow \Reals^n$ of class $C^\infty$.
\end{definition}

Let $h(\Vect{x})$ be a smooth vector field on a domain $D \in \Reals^n$. 
The \emph{Lie derivative}\index{Lie derivative} of 
$h$ along $f$, denoted as $L_f h(\Vect{x})$ can 
be defined (in local coordinates) as
\[
  \dfrac{\partial h(\Vect{x})}{\partial \Vect{x}}
  f(\Vect{x}) = \sum_{i=1}^n 
      \dfrac{\partial h(\Vect{x})}{\partial x_i}
      f(\Vect{x})
\]
since it is a smooth vector field, a Lie derivative
operator can be applied to it. Set
\[
   L_f^i = L_f L_f^{i-1}
\]

The \emph{Lie bracket}\index{Lie bracket} of $f$ 
and $g$ can be defined (in local coordinates) as
\[
  [f, g](\Vect{x}) = 
    \dfrac{\partial g}{\partial \Vect{x}} f(\Vect{x}) -
    \dfrac{\partial f}{\partial \Vect{x}} g(\Vect{x})
\]
Iterated lie brackets are denoted as $\text{ad}^i_f g$:
\begin{align*}
  \text{ad}_f g &= [f, g] \\
  \text{ad}^i_f g &= [f, \text{ad}^{i-1}_f g]
\end{align*}

Let $f_1$, \ldots, $f_\eta$ be some vector fields on
$D \subset \Reals^n$.
The \emph{distribution}\index{Distribution} 
$\Delta$ spanned the vector fields
$f_1$, \ldots, $f_\eta$ is the collection of 
vector spaces
\[
  \Delta (\Vect{x}) = \text{span}_{\Reals^n} 
    \big\{ f_1 (\Vect{x}), f_2 (\Vect{x}), \ldots, 
    f_\eta (\Vect{x}) \big\}
\]
for all $x \in D$. We denote
\[
  \Delta = \text{span}_{\Reals^n} 
    \big\{ f_1, f_2, \ldots, f_\eta \big\}
\]

A distribution $\Delta$ is \emph{involutive}\index{Distribution!Involutive} 
if 
\[
  \forall g_1, g_2 \in \Delta, [g_1, g_2] \in \Delta
\]

\begin{proposition}
The system with dyanmics 
$\dot{\Vect{x}} = f (\Vect{x}) + g (\Vect{x}) \Vect{u}$
is static state feedback linearizable if, and only if,
there is a domain $D_0 \subset D$ such that the following two conditions 
are staisfied:
\begin{enumerate}
\item The matrix $\big[ g, \text{ad}_f g, 
  \ldots, \text{ad}^{n-1}_f g  \big]$ has rank $n$
  for all $x \in D_0$.
\item The distribution $\big\{ g, \text{ad}_f g, 
  \ldots, \text{ad}^{n-2}_f g\big\}$ is involutive in
  $D_0$.
\end{enumerate}
\end{proposition}

\chapter[Definitions for extensions of differential flatness]{Precise definitions for extensions of differential flatness}
%
\section{Systems, dynamics and differential flatness}
\subsection{Basic definitions from differential algebra}
\begin{definition}
An {\em ordinary differential field} $k$, is a field on which a mapping 
$\mathsf{d}$: $k \rightarrow k$
is defined, satisfying the natural properties with respect to
addition and product, i.e., for any $x, z \in k$,
\begin{align*}
  \mathsf{d} (x + z) &= \mathsf{d}(x) + \mathsf{d}(z) \\
  \mathsf{d} (x z)   &= \mathsf{d}(x)z + x\mathsf{d}(z)
\end{align*} 
\end{definition}
\begin{definition}
Let $K$ be a field. A subfield of $K$ is a subset $k$ of $K$ that is closed under the 
field operations of $K$ and under taking inverses in $K$. In other words, $k$ is a 
field with respect to the field operations inherited from $K$. The larger field 
$K$ is then said to be an \emph{extension field} of $k$, denoted as $K/k$.
\end{definition}

\begin{definition}
Let $k$ and $K$ be differential fields with differential operators $\mathsf{d}_k$ and 
$\mathsf{d}_K$ respectively. Then, $K$ is a \emph{differential extension field} of $k$ if
$K$ is an extension field of $k$ and 
\begin{align*}
  \forall x \in k, \mathsf{d}_k (x) = \mathsf{d}_K (x)
\end{align*} 
Let $S$ be a subset of $K$. We shall denote by $k \langle S \rangle$
the differential subfield of $K$ generated by $k$ and $S$.
\end{definition}

\subsection{Algebraic and transcendental extensions} 
All fields are assumed to be of
characteristic zero. Assume also that the differential field
extension $K/k$ is {\em finitely generated}, i.e., there exists a
finite subset $S \subset K$ such that $K = k\langle S \rangle$. 
\begin{definitions}
An element $a$ of $K$ is said to be {\em differentially algebraic} over
$k$ if it satisfies an algebraic differential equation
with coefficients in $k$: there exists a non-zero polynomial $P$
over $k$, in several indeterminates, such that 
\begin{align*}
P(a, \dot{a}, \dots, a^{(\nu)}) = 0 
\end{align*}
It is said to be {\em differentially transcendental} over $k$ if it is 
not differentially algebraic. \\
The extension $K/k$ is said to be {\em differentially
algebraic} if any element of $K$ is differentially
algebraic over $k$. An extension which is not differentially
algebraic is said to be {\em differentially transcendental}. 
\end{definitions}

\subsection{Nonlinear systems and flatness}
\begin{definition}
Let $k$ be a given differential ground field. A {\em (nonlinear)
system} is a finitely generated differential
extension $K/k$.
\end{definition}
\begin{definition}
 A nonlinear system $K/k$ is called \emph{differentially flat} if there exists
 a finite family $\Vect{y} = (y_1, \ldots, y_m)$ of elements of an algebraic
 extension $L$ of $K$ such that the extension $L/k\langle \Vect{y} \rangle$ is
 (non differentially) algebraic. Such a family is called a \emph{flat output}.
\end{definition}

\section{H-fields, Liouvillian and existential closedness}
\label{app:defFlatnessExensions}
The paper
\cite{aschenbrennerEtAlNDJFL2013} reviews some of the most interesting notions for
extending the differential flatness notion. One can also see 
\cite{aschenbrennerVanDenDriesJPAA2005,aschenbrennerVanDenDriesAMS2005,vanDerHoevenSpringer2006}
for related material. 

\begin{definition}
 An $H$\emph{-field} is an ordered differential field $K$ whose natural dominance relation 
 $\preccurlyeq$ satifies the following two conditions, for all $z \in K$:
 \begin{itemize}
  \item[(H1)] If $z \succ 1$, then $\dot z / z > 0$
  \item[(H2)] If $z \preccurlyeq 1$, then $z -c \prec 1$, for some $c \in C$, the field of
    constants of $K$.
 \end{itemize}
\end{definition}
\begin{remark}
 In more usual terms, the dominance relations can be explicited as follows, for real valued
 functions:
 \begin{itemize}
  \item $f \preccurlyeq g \ \Leftrightarrow \ 
       \displaystyle{\lim_{t \rightarrow \infty}}\: \frac{f(t)}{g(t)} \in \Reals$ 
  \item $f \prec g \ \Leftrightarrow \ 
       \displaystyle{\lim_{t \rightarrow \infty}}\: \frac{f(t)}{g(t)} = 0$ 
 \end{itemize}
\end{remark}

\begin{definition}
 An $H$-field $K$ is \emph{Liouville closed} if it is real closed and any equation 
 $\dot z + az = b$ with $a, b \in K$ has a non zero solution in $K$.
\end{definition}

\begin{definition}
 An $H$-field $K$ is \emph{existentially closed} if every finite system of algebraic
 differential equations over $K$ in several unkowns with a solution in an $H$-field
 extension of $K$ has a solution in $K$.
\end{definition}

\chapter{Two link arm inverse kinematics}
\label{app:twoLinkInverseKinematics}
The position of the end effector (the wrist) of a two link arm is given by
\begin{subequations}\begin{align}
 h_x &= l_1 \cos \theta_1 + l_2 \cos (\theta_1 + \theta_2) \\
 h_y &= l_1 \sin \theta_1 + l_2 \sin (\theta_1 + \theta_2) 
\end{align}\end{subequations}
where $h_x, h_y$ are the coordinates of the end effector.
By summing the square of the two preceding equations, one obtains
\if@twocolumn
\begin{align*}
 \hspace*{-2ex}
 h_x^2 + h_y^2 &= l_1^2 + l_2^2 + 2 l_1 l_2 \big[
   \cos\theta_1 \cos (\theta_1 + \theta_2) +\\
   &\hspace*{17ex}\sin \theta_1 \sin (\theta_1 + \theta_2)
   \big] \\
  &= l_1^2 + l_2^2 + 2 l_1 l_2 \cos\theta_2
\end{align*}
\else
\begin{align*}
 h_x^2 + h_y^2 &= l_1^2 + l_2^2 + 2 l_1 l_2 \big[
   \cos\theta_1 \cos (\theta_1 + \theta_2) +
   \sin \theta_1 \sin (\theta_1 + \theta_2)
   \big] \\
  &= l_1^2 + l_2^2 + 2 l_1 l_2 \cos\theta_2
\end{align*}
\fi
Then 
\begin{align*}
 \cos \theta_2 &= \dfrac{h_x^2 + h_y^2 - l_1^2 - l_2^2}{2 l_1 l_2}
\end{align*}
or
\begin{align*}
  \theta_2 &= \arctan \left( \dfrac{\sin \theta_2}{\cos \theta_2} \right) \\
    &= \arctan \left( \dfrac{\pm \sqrt{1 - \cos^2 \theta_2}}{\cos \theta_2} \right) \\
    &= \arctan \left(
        \pm \dfrac{\sqrt{1 - \bar{h}^2}}{\bar{h}}
      \right)
\end{align*}
\begin{align*}
  \bar{h} &= \dfrac{h_x^2 + h_y^2 - l_1^2 - l_2^2}{2 l_1 l_2} \notag
\end{align*}
Setting 
\begin{align*}
 k_1 &= l_1 + l_2 \cos \theta_2, \quad k_2 = l_2 \sin \theta_2
\end{align*}
one has
\begin{align*}
  h_x &= k_1 \cos \theta_1 - k_2 \sin\theta_1 \\
  h_y &= k_1 \sin \theta_1 + k_2 \cos\theta_1 
\end{align*}
Then
\begin{align*}
  \rho &= \sqrt{k_1^2 + k_2^2}, \quad \gamma = \arctan \left( \dfrac{k_2}{k_1} \right)
\end{align*}
wherefrom
\begin{align*}
 h_x &= \rho \cos \gamma \cos \theta_1 - \rho \sin \gamma \sin \theta_1 \\
 h_y &= \rho \cos \gamma \sin \theta_1 + \rho \sin \gamma \cos \theta_1 
\end{align*}
or, what is the same
\if@twocolumn
\begin{align*}
 \dfrac{h_x}{\rho} &= \cos (\gamma + \theta_1 )\\
 \dfrac{h_y}{\rho} &= \sin (\gamma + \theta_1 )\\
\end{align*}
\else
\begin{align*}
 \dfrac{h_x}{\rho} &= \cos (\gamma + \theta_1 ),\qquad \dfrac{h_y}{\rho} = \sin (\gamma + \theta_1 )\\
\end{align*}
\fi
Then
\begin{align*}
 \theta_1 + \gamma &= \arctan \left( \dfrac{h_y}{h_x} \right)
\end{align*}
and, finally
\begin{align*}
 \theta_1 &= \arctan \left( \dfrac{h_y}{h_x} \right) - \arctan \left( 
 \dfrac{l_2 \sin \theta_2}{l_1 + l_2 \cos \theta_2}
 \right)
\end{align*}

\chapter{Two link arm code listing}
\section{Flatness based control of the arm}
%
The listing beginning on the next page is a matlab code of the differential 
flatness based control of the two link arm example whose model is depicted
in \eqref{eq:twoLinkArmModel} and tracking feedback law in \eqref{eq:twoLinkArmTrackingFeeback}.
{\onecolumn%
\begin{lstlisting}[language=matlab,style=matlab]
%%%%%%%%%%%%%%%%%%%%%%%%%%%%%%%%%%%%%%%%%%%%%%%%%%%%%%%%
% Simulation of a two joint arm flatness based tracking
%%%%%%%%%%%%%%%%%%%%%%%%%%%%%%%%%%%%%%%%%%%%%%%%%%%%%%%%
function mainTwoJointArmFlatness()
    %%%%%%%%%%%%%%%%%%%%%%%%%%%%%%%%%%%%%%%%%%%
    % Nomenclature
    % P: physical parameters
    % R: reference trajectories and conrol laws
    % U: complete control laws
    % G: gains
    % S: simulation scenario
    %%%%%%%%%%%%%%%%%%%%%%%%%%%%%%%%%%%%%%%%%%%

    clear all
    %
    %% PHYSICAL parameters
    % Fixed physical parameters
    P.g  = 9.81;     % gravity constant
    P.l1 = 0.3384;   % (m) length of lower arm part
    P.l2 = 0.4554;   % (m) length of upper arm part
    P.r1 = 0.1692;   % (m) distance to center of mass
    P.r2 = 0.2277;   % (m) distance to center of mass
    P.m1 = 2.10;     % (kg) mass of lower arm part
    P.m2 = 1.65;     % (kg) mass of upper arm part
    P.J1 = 0.025;    % (kg.m^2) inertia of lower arm part
    P.J2 = 0.075;    % (kg.m^2) inertia of upper arm part

    %% REFERENCE trajectory
    % A tanh one
    R.hxi = 0.8*(P.l1+P.l2); R.hxf = 0;
    R.hyi = P.l1 + 0.1*P.l2; R.hyf = - 0.1*P.l1;
    R.stiffness = 9; R.hxR = 0.5*R.hxi;

    %% SIMULATION values
    S.tini      = 0;     S.tend = 10;      % intial and final simulation times
    S.relTol    = 1e-3;  S.absTol = 1e-6;  % relative and absolute simul tols
    % Intial state values
    tvirt = [S.tini:0.01:S.tend]';
    strRef = inverseKinematics(tvirt, P, R, S);
    S.theta10    = strRef.theta1r(1)    + 0.05*(strRef.theta1r(end)-strRef.theta1r(1));
    S.dotTheta10 = strRef.dotTheta1r(1) + 0.05*(strRef.dotTheta1r(end)-strRef.dotTheta1r(1));
    S.theta20    = strRef.theta2r(1)    + 0.05*(strRef.theta1r(end)-strRef.theta1r(1));
    S.dotTheta20 = strRef.dotTheta2r(1) + 0.05*(strRef.dotTheta2r(end)-strRef.dotTheta2r(1));

    %% Default FEEDBACK GAINS 
    % flatness based gains
    % for (s+s1)(s+s2) = s^2 + (s1+s2)*s + s1*s2
    sTh1 = 5;  sDotTh1 = 2*sTh1;
    sTh2 = 6;  sDotTh2 = 2*sTh2;
    K.kp1 = sTh1*sDotTh1;  K.kp2 = sTh2*sDotTh2;   
    K.kd1 = sTh1+sDotTh1;  K.kd2 = sTh2+sDotTh2;
    K.Kp  = [K.kp1  0; 0  K.kp2]; 
    K.Kd  = [K.kd1  0; 0  K.kd2];

    % Saving option
    S.forSaving = 'yes';

    % Simulate    
    options = odeset('RelTol', S.relTol, 'AbsTol', S.absTol);
    [ts Xs] = ode23tb(@dynTwoJointArm, [S.tini S.tend], ...
                     [S.theta10 S.dotTheta10 S.theta20 S.dotTheta20]',...
                     options, P, R, K, S);
    S.ts = ts;  S.Xs = Xs;
    plotVariables(P, R, K, S);

end % of main()


function P = coriolisGravity(P, theta1, dotTheta1, theta2, dotTheta2)
    % Masses and intertia gathering
    m1 = P.m1; m2 = P.m2; l1 = P.l1; l2 = P.l2;
    r1 = P.r1; r2 = P.r2; J1 = P.J1; J2 = P.J2;
    % Coriolis and gravity terms computations
    M11  = J1 + J2 + (m1*r1^2) + (m2*((l1^2) + (r2^2) + (2*l1*r2*cos(theta2))));
    M12  = J2 + (m2*((r2^2) + (l1*r2*cos(theta2))));
    M21  = M12;
    M22  = J2 + (m2*r2^2);
    M  = [M11,M12; M21,M22];
    C1  = -(m2*l1*r2*dotTheta2^2*sin(theta2)) -... 
           (2*m2*l1*r2*dotTheta1*dotTheta2*sin(theta2));
    C2  = m2*l1*dotTheta1^2*r2*sin(theta2);
    C   = [C1, C2]';
    G1  = (P.g*sin(theta1)*((m2*l1)+(m1*r1))) + (P.g*m2*r2*sin(theta1+theta2));
    G2  = P.g*m2*r2*sin(theta1+theta2);
    G   = [G1, G2]';
    P.C = C;   P.G = G;  P.M = M;
end

%%%%%%%%%%%
% Dynamics
%%%%%%%%%%%
function [dotX] = dynTwoJointArm(t, X, P, R, K, S)
    persistent count;
    
    if (t <= 0) count = 0; end;
    if (round(t) >= count)
        count = count + 1;
        disp(sprintf('time t = %f\n', t));
    end;
    % Variable gathering
    dotX  = zeros(4,1);
    theta1  = X(1,:); dotTheta1  = X(2,:);  
    theta2  = X(3,:); dotTheta2  = X(4,:);  
    % Coriolis and gravity terms computations
    P = coriolisGravity(P, theta1, dotTheta1, theta2, dotTheta2);
    % Control law computation
    T = twoLinkFlatCtrlLaw(t, X, P, R, K, S);
    % Two link arm DYNAMICS
    M = P.M;  C = P.C;  G = P.G;
    ddotTheta = inv(M) * (-C - G + T);
    % return the derivative of the state
    dotX  = [dotTheta1 ddotTheta(1) dotTheta2 ddotTheta(2)]';
end

%%%%%%%%%%%%%%%%%%%%%%%%%%%%%%%%%%%%%%
% Flatness based tracking control law
%%%%%%%%%%%%%%%%%%%%%%%%%%%%%%%%%%%%%%
function [T] = twoLinkFlatCtrlLaw(t, X, P, R, K, S)    
    % Variable gathering
    dotX  = zeros(4,1);
    theta1 = X(1,:); dotTheta1 = X(2,:);  
    theta2 = X(3,:); dotTheta2 = X(4,:); 
    M = P.M;   C = P.C;  G = P.G;  l1 = P.l1;  l2 = P.l2;
    % reference trajectories
    [strRefHx strRefHy] = tanhRefTraj(t, P, R, S);
    hxr     = strRefHx.v;     hyr     = strRefHy.v;
    dotHxr  = strRefHx.d1;    dotHyr  = strRefHy.d1;
    ddotHxr = strRefHx.d2;    ddotHyr = strRefHy.d2;
    % Intermediary computations
    [Hinv phi]  = computeHinvPhi(theta1, theta2, dotTheta1, dotTheta2, P);
    hx    =  l1*cos(theta1) + l2*cos(theta1+theta2);
    hy    =  l1*sin(theta1) + l2*sin(theta1+theta2);
    dotHx = -l1*dotTheta1*sin(theta1) - l2*(dotTheta1+dotTheta2)*sin(theta1+theta2);
    dotHy =  l1*dotTheta1*cos(theta1) + l2*(dotTheta1+dotTheta2)*cos(theta1+theta2);
    % errors computation
    ehx    = hx - hxr;        ehy    = hy - hyr;
    dotEHx = dotHx - dotHxr;  dotEHy = dotHy - dotHyr;
    % Vector computation
    vectDdotHr = [ddotHxr ddotHyr]';
    vectEH     = [ehx ehy]';  
    vectDotEH  = [dotEHx dotEHy]';
    % control law computation
    T = C + G + M*Hinv*(phi + vectDdotHr -... 
                              K.Kd * vectDotEH - K.Kp * vectEH);
end

%%%%%%%%%%%%%%%%%%%%%%%%%%%%%%%%%%%
% Compute phi and the inverse of H
%%%%%%%%%%%%%%%%%%%%%%%%%%%%%%%%%%%
function [matHinv vectPhi] = computeHinvPhi(theta1, theta2, dotTheta1, dotTheta2, P)
    l1 = P.l1;  l2 = P.l2;
    hx   = l1*cos(theta1) + l2*cos(theta1+theta2);
    hy   = l1*sin(theta1) + l2*sin(theta1+theta2);
    H    = [-hy  -l2*sin(theta1+theta2); 
             hx   l2*cos(theta1+theta2)];
%    matHinv = [-l2*cos(theta1+theta2)  -l2*sin(theta1+theta2);
%                hx                      hy                   ]...
%               / (l2*(hy*cos(theta1+theta2)-hx*sin(theta1+theta2)));
    matHinv = inv(H);
    vectPhi  = [l1*dotTheta1^2*cos(theta1) + l2*(dotTheta1+dotTheta2)^2*cos(theta1+theta2);
                l1*dotTheta1^2*sin(theta1) + l2*(dotTheta1+dotTheta2)^2*sin(theta1+theta2)];
end

%%%%%%%%%%%%%%%%%%%%%%%
% Reference trajectory
%%%%%%%%%%%%%%%%%%%%%%%
function [strRefHx strRefHy] = tanhRefTraj(t, P, R, S)
  hx = ((R.hxf - R.hxi).*t )./(S.tend) + R.hxi;
  strRefHx.v = hx;
  strRefHx.d1 = ((R.hxf - R.hxi)./S.tend).*ones(length(hx),1); 
  strRefHx.d2 = zeros(length(hx),1);
  strRefTanh = tanhTr(hx, R.stiffness, R.hyi, R.hyf, R.hxR);
  strRefHy.v  = strRefTanh.v;
  strRefHy.d1 = strRefTanh.d1 .* strRefHx.d1;
  strRefHy.d2 = strRefTanh.d2 .* strRefHx.d1.^2;
end

%%%%%%%%%%%%%%%%%%%%%
% Inverse kinematics
%%%%%%%%%%%%%%%%%%%%%
function strRef = inverseKinematics(t, P, R, S)
    l1 = P.l1; l2 = P.l2;
    [strRefHx strRefHy] = tanhRefTraj(t, P, R, S);
    hxr     = strRefHx.v;     hyr  = strRefHy.v;
    dothxr  = strRefHx.d1; dothyr  = strRefHy.d1;
    ddothxr = strRefHx.d2; ddothyr = strRefHy.d2;
    cosTheta2r  = (hxr.^2 + hyr.^2 - l1.^2 - l2.^2) ./ (2.*l1.*l2);
    sinTheta2r  = sqrt(1 - cosTheta2r.^2);
    theta2r     = unwrap(atan2(sqrt(1 - cosTheta2r.^2),cosTheta2r));
    theta1r     = unwrap(atan2(hyr,hxr)) -... 
                  unwrap(atan2(l2.*sinTheta2r,(l1+l2.*cosTheta2r)));
    dotTheta2r  = -(hxr.*dothxr + hyr.*dothyr) ./ (l1*l2.*sinTheta2r);
    dotTheta1r  = (dothyr.*hxr - hyr.*dothxr) ./ (hxr.^2 + hyr.^2) -...
                   (l2.*dotTheta2r.*(1+l1.*cosTheta2r)) ./... 
                   (l1^2 + 2*l1*l2.*cosTheta2r + l2^2);
    ddotTheta2r = -(hxr.*ddothxr + dothxr.^2 + hyr.*ddothyr + dothyr.^2) ./... 
                   (l1*l2.*sinTheta2r) + (hxr.*dothxr + hyr.*dothyr) .*... 
                   (cosTheta2r.*dotTheta2r) ./ (l1*l2.*sinTheta2r.^2);
    ddotTheta1r = (ddothyr.*hxr - hyr.*ddothxr) ./ (hxr.^2 + hyr.^2) -...
                   2.*(dothyr.*hxr - hyr.*dothxr).*(hxr.*dothxr + hyr.*dothyr) -...
                   l2.*(ddotTheta2r.*(1+l1.*cosTheta2r) - (l1.*dotTheta2r.^2.*sinTheta2r)) ./...
                   (l1^2 + 2*l1*l2.*cosTheta2r + l2^2) + 2*l1*l2.*(l2.*dotTheta2r.*(1+l1 .*... 
                   cosTheta2r)).*(sinTheta2r.*dotTheta2r) ./ (l1^2+2*l1*l2.*cosTheta2r+l2^2).^2;
    strRef.theta1r     = theta1r;     strRef.theta2r     = theta2r;
    strRef.dotTheta1r  = dotTheta1r;  strRef.dotTheta2r  = dotTheta2r;
    strRef.ddotTheta1r = ddotTheta1r; strRef.ddotTheta2r = ddotTheta2r;
end

%%%%%%%%
% plots
%%%%%%%%
function plotVariables(P, R, K, S)
%
    % For plots
    police = 'Helvetica';  size = 24; lineWidth = 2;

    % reference state
    ts   = S.ts;   tr = ts;  Xs = S.Xs;
    % Reference curve and inverse kinematics
    [strRefHx strRefHy] = tanhRefTraj(tr, P, R, S);
    hxr     = strRefHx.v;     hyr     = strRefHy.v;
    dotHxr  = strRefHx.d1;    dotHyr  = strRefHy.d1;
    ddotHxr = strRefHx.d2;    ddotHyr = strRefHy.d2;
    strRefTh    = inverseKinematics(tr, P, R, S);
    theta1r     = strRefTh.theta1r;     theta2r     = strRefTh.theta2r;
    dotTheta1r  = strRefTh.dotTheta1r;  dotTheta2r  = strRefTh.dotTheta2r;
    % simulated state
    theta1 = Xs(:,1);  dotTheta1 = Xs(:,2); 
    theta2 = Xs(:,3);  dotTheta2 = Xs(:,4); 
    % Control laws computation
    T1r = []; T2r = [];  T1 = []; T2 = [];
    XsT = Xs';
    for i = 1:length(ts)
        % open loop control
        Pr = coriolisGravity(P, theta1r(i), dotTheta1r(i), theta2r(i), dotTheta2r(i));
        Mr = Pr.M;   Cr = Pr.C;  Gr = Pr.G;
        [Hinvr phir] = computeHinvPhi(theta1r(i), theta2r(i), dotTheta1r(i), dotTheta2r(i), P);
        DdotHr = [ddotHxr(i) ddotHyr(i)]';
        Tr = Cr + Gr - Mr*Hinvr*(DdotHr + phir);
        T1r = [T1r; Tr(1)];  T2r = [T2r; Tr(2)];
        % simulated closed loop control 
        P = coriolisGravity(P, theta1(i), dotTheta1(i), theta2(i), dotTheta2(i));
        T = twoLinkFlatCtrlLaw(ts(i), XsT(:,i), P, R, K, S);
        T1  = [T1; T(1)]; T2  = [T2; T(2)];
    end;
    % Simulated curve: forward kinematics
    l1 = P.l1;   l2 = P.l2;
    hx     = l1.*cos(theta1) + l2 .* cos(theta1 + theta2);
    hy     = l1.*sin(theta1) + l2 .* sin(theta1 + theta2);
    
    figure(1);
    % hx hy plot
    subplot(2, 2, 1); 
    plot(tr, hxr, 'r', ts, hx, 'b', 'LineWidth', lineWidth);  grid;
    xlabel('time (s)', 'FontName', police, 'FontSize', size);
    ylabel('h_x, h_{xr} (deg)', 'FontName', police, 'FontSize', size);
    title('red h_{xr} ; blue h_x',...
          'FontName', police, 'FontSize', size, 'FontWeight','bold');
    % theta2 plot
    subplot(2, 2, 2);
    plot(tr, hyr, 'r', ts, hy, 'b', 'LineWidth', lineWidth);  grid;
    xlabel('time (s)', 'FontName', police, 'FontSize', size);
    ylabel('h_y, h_{yr} (deg)', 'FontName', police, 'FontSize', size);
    title('red h_{yr} ; blue h_y',...
          'FontName', police, 'FontSize', size, 'FontWeight','bold');
    subplot(2, 2, 3);
    plot(tr, theta1r.*(180/pi), 'r', ts, theta1.*(180/pi), 'b', 'LineWidth', lineWidth);  grid;
    xlabel('time (s)', 'FontName', police, 'FontSize', size);
    ylabel('\theta_{2}, \theta_{2r} (deg)', 'FontName', police, 'FontSize', size);
    title('red \theta_{1r} ; blue \theta_1',...
          'FontName', police, 'FontSize', size, 'FontWeight','bold');
    subplot(2, 2, 4);
    plot(tr, theta2r.*(180/pi), 'r', ts, theta2.*(180/pi), 'b', 'LineWidth', lineWidth);  grid;
    xlabel('time (s)', 'FontName', police, 'FontSize', size);
    ylabel('\theta_{2}, \theta_{2r} (deg)', 'FontName', police, 'FontSize', size);
    title('red \theta_{2r} ; blue \theta_2',...
          'FontName', police, 'FontSize', size, 'FontWeight','bold');

    hFig2 = figure(2);
    % T1 and T1r plot
    set(gca, 'FontName', police, 'FontSize', size);
    plot(tr, T1r, 'r--', ts, T1, 'b-', 'LineWidth', lineWidth);  grid;
    xlabel('time (s)', 'FontName', police, 'FontSize', size);
    ylabel('T_1, T_{1r} (N)', 'FontName', police, 'FontSize', size);
    if (strcmp(S.forSaving, 'yes') ~= 1)
        title('red T_{1r} ; blue T_1',...
          'FontName', police, 'FontSize', size, 'FontWeight','bold');
    else
        print(hFig2, '-dpdf', '../GraphicsImages/twoLinkArmCtrl1.pdf');
    end

    hFig3 = figure(3);
    % T2 and T2r plot
    set(gca, 'FontName', police, 'FontSize', size);
    plot(tr, T2r, 'r--', ts, T2, 'b-', 'LineWidth', lineWidth);  grid;
    xlabel('time (s)', 'FontName', police, 'FontSize', size);
    ylabel('T_2, T_{2r} (N)', 'FontName', police, 'FontSize', size);
    if (strcmp(S.forSaving, 'yes') ~= 1)
        title('red T_{2r} ; blue T_2',...
               'FontName', police, 'FontSize', size, 'FontWeight','bold');
    else
        print(hFig3, '-dpdf', '../GraphicsImages/twoLinkArmCtrl2.pdf');
    end

    hFig4 = figure(4);
    set(gca, 'FontName', police, 'FontSize', size);
    plot(tr, hx-hxr, 'b', 'LineWidth', lineWidth);  grid;
    xlabel('time (s)', 'FontName', police, 'FontSize', size);
    ylabel('h_x - h_{xr} (m)', 'FontName', police, 'FontSize', size);
    if (strcmp(S.forSaving, 'yes') ~= 1)
        title('h_x - h_{xr}',...
              'FontName', police, 'FontSize', size, 'FontWeight','bold');
    else
        print(hFig4, '-dpdf', '../GraphicsImages/twoLinkArmErrHx.pdf');
    end

    hFig5 = figure(5);
    set(gca, 'FontName', police, 'FontSize', size);
    plot(tr, hy-hyr, 'b', 'LineWidth', lineWidth);  grid;
    xlabel('time (s)', 'FontName', police, 'FontSize', size);
    ylabel('h_y - h_{yr} (m)', 'FontName', police, 'FontSize', size);
    if (strcmp(S.forSaving, 'yes') ~= 1)
        title('h_y - h_{yr}',...
              'FontName', police, 'FontSize', size, 'FontWeight','bold');
    else
        print(hFig5, '-dpdf', '../GraphicsImages/twoLinkArmErrHy.pdf');
    end

    % forward reference and actual kinematics
    hFig6 = figure(6);
    set(gca, 'FontName', police, 'FontSize', size);
    plot(hx, hy, 'b-', hxr, hyr, 'r--', 'LineWidth', lineWidth);  grid;
    xlabel('h_x (m)', 'FontName', police, 'FontSize', size);
    ylabel('h_y (m)', 'FontName', police, 'FontSize', size);
    if (strcmp(S.forSaving, 'yes') ~= 1)
	title('blue h_x h_y  red h_{xr} h_{yr}',...
          'FontName', police, 'FontSize', size, 'FontWeight','bold');
    else
        print(hFig6, '-dpdf', '../GraphicsImages/twoLinkArmHxHy.pdf');
    end

    hFig7 = figure(7);
    set(gca, 'FontName', police, 'FontSize', size);
    R.hxi = 0.8*(P.l1+P.l2); R.hxf = 0;
    R.hyi = P.l1 + 0.1*P.l2; R.hyf = - 0.1*P.l1;
    plot([], []); grid;
    hold on    
    for i = 1:3:length(ts)
        line([0, l1*cos(theta1(i))], [0, l1*sin(theta1(i))], 'Color', [0 0 0.5], 'LineWidth', 2);
        line([l1*cos(theta1(i)), l1*cos(theta1(i))+l2*cos(theta1(i)+theta2(i))],...
             [l1*sin(theta1(i)),...
              l1*sin(theta1(i))+l2*sin(theta1(i)+theta2(i))], 'Color', [0 0 0.9], 'LineWidth', 2);
        plot(hx(i), hy(i), 'ro', 'LineWidth', 2); grid;
        pause(0.01);
    end
    hold off
    xlabel('h_x (m)', 'FontName', police, 'FontSize', size);
    ylabel('h_y (m)', 'FontName', police, 'FontSize', size);
    if (strcmp(S.forSaving, 'yes') ~= 1)
        title('blue h_x h_y  red h_{xr} h_{yr}',...
          'FontName', police, 'FontSize', size, 'FontWeight','bold');
    else
        print(hFig7, '-dpdf', '../GraphicsImages/twoLinkArmAnimation.pdf');
    end    

end

\end{lstlisting}%
}%

\end{appendix}

\end{document}